\documentclass[12pt,a4paper]{article}

\RequirePackage[OT1]{fontenc}
\RequirePackage{amsthm}
\RequirePackage[cmex10]{amsmath}
\RequirePackage{natbib}
\RequirePackage[colorlinks,citecolor=blue,urlcolor=blue, linkcolor=blue, bookmarksopen=true]{hyperref}

\usepackage[margin=1in]{geometry}
\usepackage[latin1]{inputenc}
\usepackage{amsmath}
\usepackage{amsfonts}
\usepackage{amssymb}
\usepackage{comment}
\usepackage{bbm}
\usepackage{graphicx}
\usepackage[usenames,dvipsnames]{color}
\usepackage{multirow}
\usepackage[labelfont=bf]{caption}
\usepackage{placeins}
\usepackage{array}
\usepackage{enumitem}
\usepackage{rotating}
\usepackage{adjustbox}
\usepackage{mathrsfs}
\usepackage{gensymb}
\usepackage{float}
\usepackage{subcaption}
\usepackage{wasysym}
\usepackage{siunitx}

\newcolumntype{R}[2]{%
    >{\adjustbox{angle=#1,lap=\width-(#2)}\bgroup}%
    l%
    <{\egroup}%
}

\title{\textsc{Subjective inference}\protect}
\renewcommand\footnotemark{}

\author{\textsc{Andrew Mackenzie}* \thanks{*Department of Economics, Rutgers University. Contact email: [andrew.k.mackenzie@rutgers.edu].}}
\date{First draft: October 31, 2025 \\ This draft: \today}

\begin{document}

\maketitle

\begin{abstract}
An agent observes a {\it clue}, and an analyst observes an {\it inference}: a ranking of events on the basis of how corroborated they are by the clue. We prove that if the inference satisfies the axioms of \cite{Villegas1964} except for the classic qualitative probability axiom of monotonicity, then it has a unique normalized signed measure representation (\hyperlink{Theorem1}{Theorem~1}). Moreover, if the inference also declares the largest event equivalent to the smallest event, then it can be represented as a difference between a posterior and a prior such that the former is the conditional probability of the latter with respect to an assessed event that is interpreted as a clue guess. Across these {\it Bayesian representations}, the posterior is unique, all guesses are in a suitable sense equivalent, and the prior is determined by the weight it assigns to each possible guess (\hyperlink{Theorem2}{Theorem~2}). However, observation of a prior and posterior compatible with the inference could reveal that all of these guesses are wrong.\\
\end{abstract}

\hypertarget{Section1}{}
\section{Introduction}

\hypertarget{Section1.1}{}
\subsection{Overview}

In this paper, we consider situations where an {\it agent} privately observes an abstract {\it clue} and an {\it analyst} observes an assessment of some events that the agent makes on the basis of this private information. In the classic framework for revealing subjective beliefs through behavior (\citealp{Ramsey1931}; \citealp{Savage1954}), the agent assesses events on the basis of relative likelihood, which the analyst observes if (i)~each event has a security that pays $\$1$ if the event occurs and $\$0$ otherwise, (ii)~the agent would rather surely receive $\$1$ than $\$0$, and (iii)~the analyst observes how the agent ranks securities {\it when they are offered for free}. We have in mind that instead of observing the prior belief or the posterior belief, the analyst instead observes an {\it inference} ranking events on the basis of how corroborated they are by the clue, which intuitively captures the direction of belief revision without revealing its origin, destination, or magnitude. More concretely, if (i)~the agent is risk neutral, (ii)~the agent's prior matches the current prices in the market for securities, and (iii)~the analyst observes the agent's posterior ranking of securities {\it when they are offered at current prices}, then the analyst observes what we call an inference even without observing the prices themselves. In this case, what can the analyst infer from the agent's behavior?

It can, of course, be beneficial to make an inference about somebody else's private information. As a simple example, suppose that the analyst observes the following: the agent steps outside, looks at his phone for a few moments, raises his eyebrows, returns inside, and then reappears with an umbrella. Though the analyst does not know what the agent observed on his phone, the agent's behavior before and after he acquires this clue allow the analyst to make an inference about the agent's inference, and this could benefit an analyst who prefers to stay dry. As a second example, consider an analyst who stands in for ``the price system" in a securities market. The efficient market hypothesis posits that security prices ``fully reflect" all available information \citep{Fama1970}, and in order for this to be true in a market where messages express demand, it must be the case that after an agent observes an informative clue---even a clue that does not happen to have its own associated security---some ``relevant information" must be determined by the agent's posterior ranking of the securities at current prices.

We formalize inferences using the classic model of qualitative probabilities (\citealp{Bernstein1917}; \citealp{deFinetti1937}; \citealp{Koopman1940}; \citealp{Savage1954}; \citealp{Villegas1964}): there is a $\sigma$-algebra of events, each of which may be interpreted as a logical proposition that is either true or false, and these are ranked on the basis of some assessment. The classic interpretation that these events are ranked on the basis of relative likelihood is captured in part by the monotonicity axiom, which requires that $A \subseteq B$ implies $B \succsim A$, and this is a compelling normative requirement when an agent compares securities that are offered for free. That said, monotonicity is not compelling when an agent compares securities that are offered at current prices: an informative clue may well reveal an underpriced security that can be purchased profitably, but the security that offers $\$1$ surely should always be priced at $\$1$ and hence never be profitable to purchase.

Motivated by this reasoning, we define an {\it inference} to be a complete and transitive ranking of events that satisfies separability: if three events $A$, $B$, and $C$ are such that (i)~$A$ and $C$ are disjoint, and (ii)~$B$ and $C$ are disjoint, then $A \succsim B$ if and only if $A \cup C \succsim B \cup C$. This allows for the classic interpretation that events are compared on the basis of relative likelihood, but also for the interpretation that they are compared on the basis of {\it change} in likelihood: $A \succ B$ means that $A$ is {\it more corroborated by the clue than} $B$, in the sense that the clue's absolute impact on how much more likely the agent finds an event is greater for $A$ than $B$. Indeed, the same sort of contingent reasoning that motivates separability for relative likelihood and the sure-thing principle for preferences \citep{Savage1954} also motivates separability for relative corroboration. Notice that this description of relative corroboration makes no reference to a security market, and that is because the notion does not actually require one; securities and their prices simply provide one concrete story for how an inference might be observed in practice.

With relative likelihood, no event is ever ranked below the impossible event, but with relative corroboration, there will generally be some negatively corroborated events. For example, a detective might find a shoeprint exonerating for the maid and incriminating for the butler. This motivates expanding the set of representations we consider from probability measures to {\it signed measures}, which can assign real numbers outside of the unit interval. Signed measures have previously appeared in the decision theory literature, notably including the recent paper of \cite{Brandenburger-Ghirardato-Pennesi-Stanca2024} preceding this paper that generalizes \cite{Anscombe-Aumann1963}; see their paper for other references. We deviate from these papers by focusing entirely on the subjective assessment of events.

In particular, the classic theorem of \cite{Villegas1964} states that the four qualitative probability axioms (including monotonicity), a continuity axiom, and an axiom that forbids atoms together imply that there is a unique probability measure representation. We prove that if monotonicity is dropped from the list of hypotheses, then there is a unique normalized signed measure representation (\hyperlink{Theorem1}{Theorem~1}). Intuitively, this can be interpreted as a {\it direction} of belief revision when the $\sigma$-algebra of assessed events does not necessarily include the sure event. More precisely, if the collection of all events including the sure event is $\mathcal{A}^*$, the prior is $\mu_0$, the posterior is $\mu_1$, and the inference ranks events in the $\sigma$-algebra $\mathcal{A} \subseteq \mathcal{A}^*$ according to the belief revision $\mu_1 - \mu_0$, then the signed measure representation of the inference is the restriction of $\mu_1 - \mu_0$ to $\mathcal{A}$ with its magnitude erased through normalization.

Intuitively, if the $\sigma$-algebra of assessed events includes the sure event, then its largest event remains sure and its smallest event remains impossible after the clue is acquired. In this spirit, if in addition to the hypotheses of the previous theorem, the inference declares the largest event equivalent to the smallest event, then it can be represented as a difference between a posterior and a prior, where the former is the conditional probability of the latter with respect to an assessed event that is interpreted as a clue guess. Across these {\it Bayesian representations}, the posterior is unique, all guesses are in a suitable sense equivalent, and the prior is determined by the weight it assigns to each possible guess (\hyperlink{Theorem2}{Theorem~2}). We caution, however, that even though such an inference never falsifies the hypothesis that the clue was an assessed event, and therefore allows the analyst to guess that the clue was an assessed event, that does not make the hypothesis true. Indeed, a prior and posterior compatible with the inference {\it can} falsify the hypothesis that the clue was an assessed event, and we illustrate this with a concrete example in the next section.

\hypertarget{Section1.2}{}
\subsection{Illustrative example: the drunk archer}

In this section, we discuss the interpretation of an inference and its representations in the context of a concrete example.

\vspace{\baselineskip} \noindent \textsc{The story.} We have in mind a future scheduled event with an active betting market, such as an election or a sports game, for which some information might be acquired between now and the event, such as polling data or injury reports. For the sake of concreteness, we take as our example an archery competition.

In particular, suppose there is an active and sophisticated betting market dedicated to the first arrow of the reigning champion that strikes the target, suppose the target is a circle, and suppose it is certain the champion will hit the target at least once over the course of the competition. We model the associated set of {\it arrow states} as the unit circle, $S = \{(x, y) \in \mathbb{R}^2 | x^2 + y^2 \leq 1\}$, and the collection of available {\it arrow securities} as the $\sigma$-algebra of Lebesgue measurable subsets of $S$, $\mathcal{A} \subseteq 2^S$. For example, $\{ (x, y) \in S | x \geq 0 \}$ represents the security that pays $\$1$ if the arrow hits the right side of the target and $\$0$ otherwise.\footnote{These arrow securities for events in our infinite-state model play the role of Arrow securities for states in finite-state models \citep{Arrow1964}.} Assume the analyst knows that, at least initially, the agent believes all events with zero Lebesgue measure are null.

Our agent engages enthusiastically with this security market in the days leading up to the competition, and the analyst observes the agent's behavior. The morning of the competition, the agent suddenly reveals a new interest in securities for which the arrow misses the bullseye. The analyst knows that the agent has privately observed a clue, but does not know what it was. What can the analyst infer?

\vspace{\baselineskip} \noindent \textsc{The assessments.} In our story, the agent makes three assessments: the {\it prior belief} ranks securities when they are offered for free before acquiring the clue, the {\it posterior belief} ranks securities when they are offered for free after acquiring the clue, and the {\it inference} ranks securities when they are offered for current market prices after acquiring the clue. We consider two scenarios---one where the clue has an associated security and one where it does not---and use four measures to make these scenarios concrete.

In particular, let $B \equiv \{(x, y) \in S | x^2 + y^2 \leq \frac{1}{2}\}$ denote the {\it bullseye}. The four measures vary in how much probability they assign to the bullseye; we refer to them as very high measure $\mu_{VH}$, high measure $\mu_H$, low measure $\mu_L$, and very low measure $\mu_{VL}$. To define them, first let $\mu_{\ell}: \mathcal{A} \to [0, 1]$ denote the Lebesgue measure, which maps each region to its area and intuitively plays the role of our uniform distribution. Moreover, let $\mu_B$ denote the conditional probability of $\mu_{\ell}$ given $B$, and let $\mu_{S \backslash B}$ denote the conditional probability of $\mu_{\ell}$ given $S \backslash B$; thus for each $A \in \mathcal{A}$ we have $\mu_B(A) = \frac{\mu_{\ell}(A \cap B)}{\mu_{\ell}(B)} = 4 \cdot \mu_{\ell}(A \cap B)$ and $\mu_{S \backslash B}(A) = \frac{\mu_{\ell}(A \cap B)}{\mu_{\ell}(S \backslash B)} = \frac{4}{3} \cdot \mu_\ell(A \backslash B)$. We can now define the four measures in our story: for each $A \in \mathcal{A}$, we define
\begin{itemize}
\item $\mu_{VH}(A) \equiv \frac{9}{10} \cdot \mu_B(A) + \frac{1}{10} \cdot \mu_{S \backslash B}(A)$,

\item $\mu_{H}(A) \equiv \frac{77}{100} \cdot \mu_B(A) + \frac{23}{100} \cdot \mu_{S \backslash B}(A)$,

\item $\mu_{L}(A) \equiv  \frac{1}{4} \cdot \mu_B(A) + \frac{3}{4} \cdot \mu_{S \backslash B}(A)$, which is equal to $\mu_{\ell}(A)$, and

\item $\mu_{VL}(A) \equiv \mu_{S \backslash B}(A)$.
\end{itemize}
Observe that the probability that the arrow strikes the bullseye is 90\% according to $\mu_{VH}$, 77\% according to $\mu_H$, 25\% according to $\mu_L$, and 0\% according to $\mu_{VL}$.

Throughout our discussion, we assume that the prior belief is represented by $\mu_H$, and moreover we assume that this matches the prices in the security market. We make the latter assumption as a storytelling device, to provide a simple and concrete interpretation for the inference that the analyst observes, and caution that this is rather restrictive: even in a world with objective probabilities, state prices generally deviate from them to adjust for time preferences and risk preferences (see for example \citealp{Dybvig-Ross2003}). In the same spirit, we also assume that the agent is risk neutral.

\vspace{\baselineskip} \noindent \textsc{Scenario 1.} It is instructive to first analyze an unrealistic but simple scenario: the analyst is sure that the abstract clue is in fact an assessed event in $\mathcal{A}$, or that the agent has observed a region that the arrow surely struck. For example, perhaps the contest has already been pre-recorded but not yet broadcast, and the agent has a friend involved in the recording who privately leaks some information to him.

To be concrete, suppose that the prior is represented by the high measure $\mu_H$, the clue is that the arrow missed the bullseye modeled as the event $S \backslash B$, and the posterior is derived from the prior via Bayesian updating. In this case, the posterior is represented by the conditional probability of $\mu_H$ given $S \backslash B$, which assigns to each event $A$ the probability $\frac{\mu_H(A \cap (S \backslash B))}{\mu_H(S \backslash B)}$, and this is exactly the very low measure $\mu_{VL}$. It is not hard to see that both the prior and the posterior have unique measure representations; thus if the analyst observes one of these rankings, then he effectively observes the associated measure. Of course, what the analyst can infer about the clue depends on what behavior of the agent he observes.

If the analyst is fortunate enough to observe both the prior and the posterior, then he can infer that the clue is approximately $S \backslash B$, where formally we say that (i)~an event is purely null if all of its subevents are null, and (ii)~two events $A$ and $A'$ are approximately equivalent if their symmetric difference $(A \cup A') \backslash (A \cap A')$ is purely null according to the inference. This distinction between approximate and precise identification of the clue is a minor technical point unrelated to the main ideas of this discussion; the main point here is that with both the prior and the posterior, the analyst essentially infers the clue. Indeed, the analyst effectively observes both $\mu_H$ and $\mu_{VL}$, and therefore infers that the clue is an event that maximizes $\mu_{VL} - \mu_H$.

What if the analyst only observes one of these rankings? If he only observes the prior, then of course he can infer nothing: this ranking does not depend in any way on the clue. On the other hand, if he only observes the posterior, then he can infer something, although less than he could infer with both the prior and the posterior: he infers that the clue is approximately an event that contains $S \backslash B$. Indeed, the analyst effectively observes $\mu_{VL}$, and therefore infers that the clue does not rule out any event outside of the bullseye that has positive Lebesgue measure.

Finally, what if again the analyst only observes one ranking, but this time it is the agent's {\it inference} ranking securities when they are offered at current prices after the agent acquires the clue? Because we have assumed that the agent is risk neutral and the current prices match the prior $\mu_H$, this is represented by $\mu_{VL} - \mu_H$: for each $A \in \mathcal{A}$, the agent assesses the expected utility of purchasing the security for $A$ after acquiring the clue to be $\mu_{VL}(A) \cdot (1 - \mu_H(A)) + (1-\mu_{VL}(A)) \cdot (-\mu_H(A)) = \mu_{VL}(A) - \mu_H(A)$, which for brevity we refer to as the {\it score of $A$}. Because the prior matches the market prices, an event's score quantifies both the agent's assessment of how corroborated it is by the clue and the agent's assessment of its profitability at current prices. Notice that the analyst does {\it not} effectively observe these scores, because the observed ranking can also be represented as a belief difference using, for example, the prior $\frac{2}{3} \cdot \mu_H + \frac{1}{3} \cdot \mu_{VL}$ and the same posterior $\mu_{VL}$. Even so, the analyst can infer that the clue is among the top-ranked events, and therefore again draw the stronger conclusion that it is approximately $S \backslash B$.

In each Bayesian representation of the inference, the analyst guesses that the clue is something equivalent to $S \backslash B$ and correctly identifies the posterior as $\mu_{VL}$. Indeed, the analyst reasons that each negative event in the inference should be null in the posterior and the positive events should be ranked the same way by the inference and the posterior, and in doing so he correctly infers the posterior. As for the prior, the analyst is not sure: the inference reveals that the prior is something in the direction of $\mu_H$ from $\mu_{VL}$, but it does not reveal the magnitude of the belief revision.

\vspace{\baselineskip} \noindent \textsc{Scenario 2.} Suppose now that the prior is $\mu_H$ and the posterior is $\mu_L$. These measures are distinct and yet they agree on the set of null securities, so there is no $A \in \mathcal{A}$ such that $\mu_L$ is the conditional probability of $\mu_H$ given $A$. That said, the agent may yet prove to be Bayesian after all: $\mathcal{A}$ is the set of events for which securities are available, but there may be other events because markets may be incomplete.

Indeed, suppose that the morning of the competition, the agent privately observes that the reigning champion is drunk. Because this clue is not a Lebesgue measurable subset of the unit circle, it does not belong to $\mathcal{A}$, so the sophisticated betting market dedicated to the arrow's location does not offer a security for this event, and thus the analyst does not observe how the agent assesses the likelihood of this event. One of the standard reasons for incomplete markets is that it is difficult to enforce a contract in court when it is difficult to establish which of its clauses hold, and the security that pays $\$1$ when the champion is drunk is a natural example of a contract for which this is a concern: the champion may well refuse any test that a court would require. Even so, the agent in our story knows what he sees.

In order to model this situation in a manner that captures all relevant uncertainty including the agent's clue, we must extend our previous model. To do so, let us define the {\it extended state space} $S^* \equiv S \times \{s, d\}$, where $s$ denotes that the champion is sober and $d$ denotes that the champion is drunk, and let $\mathcal{A}^*$ denote the associated product $\sigma$-algebra for $\mathcal{A}$ and the power set of $\{s, d\}$. We refer to members of $\mathcal{A}^*$ as {\it events}. Abusing notation, we identify each arrow security $A$ with the event $A \times \{s, d\}$ and view $\mathcal{A}$ as a subset of $\mathcal{A}^*$. Crucially, the clue that the champion is drunk is an example of an event that is not a security.

Suppose the agent's {\it extended prior} $\mu: \mathcal{A}^* \to [0, 1]$ is such that (i)~the probability the champion is sober is $\frac{4}{5}$, (ii)~the conditional probability on $\mathcal{A}$ given that the champion is sober is $\mu_{VH}$, (iii)~the probability the champion is drunk is $\frac{1}{5}$, and (iv)~the conditional probability on $\mathcal{A}$ given that the champion is drunk is $\mu_L$. Then the observed prior, or the restriction of the extended prior to $\mathcal{A}$, is $\mu_H$. Moreover, if the agent observes that the champion is drunk and then revises the extended prior in accordance with Bayesian updating to form the {\it extended posterior}, then the observed posterior, or the restriction of the extended posterior to $\mathcal{A}$, is $\mu_L$. The agent is indeed Bayesian after all!

Within the model $(S, \mathcal{A})$, the revision from the prior $\mu_H$ to posterior $\mu_L$ involves shifting weight from the bullseye to its complement, but not until no weight on the bullseye remains. This revision is a generalization of Bayesian updating known as Jeffrey conditionalization \citep{Jeffrey1965}, which can also be interpreted using a {\it different} extended model as the agent's response to a noisy signal about whether or not the arrow misses the bullseye; here the extended states are used to specify signal realizations conditional on the arrow state. The analyst does not know which extended model describes the truth, but in some sense it does not matter: in any case, $S \backslash B$ is an example of an event in the restricted model that is most corroborated by the agent's clue, and as in the previous scenario, what the analyst can infer about $S \backslash B$ depends on what behavior of the agent he observes.

This scenario's analysis of what the analyst can infer about the securities that are most corroborated by the clue is largely unchanged from the previous scenario, with one exception: if the analyst only observes the posterior, then he infers that either the clue is equivalent to $S$ or the clue has no associated security, because the clue does not rule out {\it any} region of the target with positive Lebesgue measure. Even so, the analyst can still infer from the agent's inference that all top-ranked arrow securities are approximately equivalent to $S \backslash B$.

The Bayesian representations of the inference are identical to those in the previous scenario for exactly the same reasons. In this scenario, however, the analyst's guesses are wrong. Indeed, the analyst again reasons that each negative event in the inference should be null in the posterior, and this reasoning is based on the assumption that the clue is an event that is assessed in the inference, but while this assumption was correct in the previous scenario, it is incorrect in this one: the clue is an event without a security, and thus it is not assessed by the inference. Altogether, then, under the assumption that the agent revises his beliefs in accordance with Bayesian updating,\footnote{Of course, this assumption can be challenged. For a recent overview of the criticisms of Bayesian updating as well as an alternative, see \cite{Dominiak-Kovach-Tserenjigmid2025}.} the prior and posterior can together falsify the hypothesis that the clue has an associated security, but the inference alone cannot.

\vspace{\baselineskip} \noindent \textsc{Discussion.} We conclude our discussion of this archery example with three remarks.

First, in both scenarios, the analyst can infer that the agent's favorite securities are approximately equivalent to $S \backslash B$ without observing the agent's full inference; it suffices to observe one of the agent's favorite securities. That said, the inference and its normalized signed measure representation provide more information, and this additional information can be useful. For example, if the analyst would like to revise security prices after observing the agent, then observation that one of the agent's favorite securities is $A$ leaves many options, the simplest of which involve fixing one proportion by which to increase the price of each subevent of $A$ and fixing a second proportion by which to decrease the price of each subevent of $S \backslash A$, such that the prices of other securities are determined through additivity and the price of $S$ is unchanged. In a more complex example, however, this simple policy could {\it create} new profitable opportunities for the agent, and the analyst could avoid this by observing the entire inference and adjusting prices in proportion to the normalized signed measure.

Second, for each assessment in our example, the collection of top-ranked events was nonempty. What about in general? This is of course a trivial point for beliefs, as monotonicity implies that the largest event is top-ranked, but it is a surprisingly delicate point for inferences: even under the additional assumptions that the ranking is monotonic and that there is a unique null event, the natural topology is not compact. Indeed, under these additional assumptions, (i)~by \cite{Villegas1964}, the $\sigma$-algebra carries a countably additive measure $\mu$ that is strictly positive, and (ii)~by \cite{Mackenzie2019}, {\it monotone continuity} is equivalent to the requirement that each upper contour set and each lower contour set is closed in the topology of sequential order convergence, $\tau_{soc}$.\footnote{An event sequence $(A_i)$ {\it order-converges} if its limit superior $\cap_{i \in \mathbb{N}} \cup_{j \geq i} A_j$ and limit inferior $\cup_{i \in \mathbb{N}} \cap_{j \geq i} A_j$ coincide, in which case this event is its {\it order-limit} $\lim A_i$. The {\it topology of sequential order convergence} is the finest topology such that each closed set includes the order-limit of every sequence of its points that order-converges. For further reading, see \cite{Maharam1947}, \cite{Vladimirov2002} and \cite{Balcar-Jech-Pazak2005}.} If $\tau_{soc}$ were compact, then we could immediately conclude that the intersection of all upper contour sets is nonempty in order to establish that there is a top-ranked event. But (i)~since the $\sigma$-algebra carries a strictly positive measure, thus $\tau_{soc}$ is metrizable \citep{Maharam1947}, so $\tau_{soc}$ is compact if and only if it is sequentially compact, and (ii)~since $\mu$ is countable additive, it is continuous, so it is a strictly positive Maharam submeasure on $\mathcal{A}'$, so $\mathcal{A}'$ {\it adds independent reals} (\citealp{Balcar-Jech-Pazak2005}; \citealp{Velickov2005}), so $\tau_{soc}$ is not sequentially compact \citep{Balcar-Jech-Pazak2005}. Altogether, then, $\tau_{soc}$ is not compact, so we must establish that there is a top-ranked event some other way. Our signed measure representation addresses this deep technical issue because it has a {\it Hahn decomposition}: a disjoint pair of events $(H^\oplus, H^\ominus)$ whose union is $S$ such that $H^\oplus$ only contains events with non-negative scores and $H^\ominus$ only contains events with non-positive scores. Though there is not a unique Hahn decomposition, all positive events that appear in Hahn decompositions are approximately equivalent and top-ranked by the inference.

Finally, in our discussion, the analyst observed complete rankings for the restricted model $(S, \mathcal{A})$ without understanding which extension of that model represented the truth. The analysis is identical, however, if the analyst knows the complete model $(S^*, \mathcal{A}^*)$ but only observes an incomplete ranking of $\mathcal{A}^*$ with the special feature that it completely ranks all members of a $\sigma$-algebra $\mathcal{A} \subseteq \mathcal{A}^*$. This is compatible with our previous incomplete market explanation, because even though the analyst understands the extended model he may still only observe how the agent ranks securities. That said, it is also compatible with complete markets, where it is possible to bet on whether or not the champion is drunk and where the analyst observes everything, if instead the source of incompleteness in the observed ranking is indecisiveness in beliefs (\citealp{Bewley2002}; \citealp{Ok-Ortoleva-Riella2012}). Regardless of the source of incompleteness, the perspective that the inference completely ranks events in a $\sigma$-algebra $\mathcal{A}$ contained in the collection of all events $\mathcal{A}^*$ is useful, because it allows for the possibility that $\mathcal{A}$ does not include the sure event, and intuitively this is the possibility allowed by \hyperlink{Theorem1}{Theorem~1} but forbidden by \hyperlink{Theorem2}{Theorem~2}.

\hypertarget{Section2}{}
\section{Model}

\hypertarget{Section2.1}{}
\subsection{Axioms for event comparisons}

An individual assesses a collection of events. We assert that the collection of events forms a Boolean algebra, allowing for the interpretation that events are logical propositions that can be combined using conjunction ({\it and}), disjunction ({\it or}), and negation ({\it not}). Moreover, we assert that the collection of events forms a $\sigma$-algebra, which makes our model consistent with classic models of probability while powering the continuity axiom that is central to our analysis. That said, we deviate from the standard approach in decision theory in that we do not require there to be a state space such that each event is a collection of states.

\vspace{\baselineskip} \noindent \textsc{Definition:} A partially ordered set $(\mathcal{A}, \supseteq)$ is a {\it lattice} if and only if for each pair $A, B \in \mathcal{A}$, there is a supremum $A \cup B$ and an infimum $A \cap B$. A lattice $(\mathcal{A}, \supseteq)$ is a {\it $\sigma$-algebra} if and only if
\begin{itemize}
\item it is {\it distributive}: for each triple $A, B, C \in \mathcal{A}$, $A \cap (B \cup C) = (A \cap B) \cup (A \cap C)$ and $A \cup (B \cap C) = (A \cup B) \cap (A \cup C)$;

\item it is {\it complemented}: there are $S, \emptyset \in \mathcal{A}$ such that for each $A \in \mathcal{A}$, $S \supseteq A \supseteq \emptyset$, and moreover, for each $A \in \mathcal{A}$, there is a unique $\neg A \in \mathcal{A}$ such that $A \cup \neg A = S$ and $A \cap \neg A = \emptyset$; and

\item each countably infinite collection $\{ A_i \}_{i \in \mathbb{N}} \subseteq \mathcal{A}$ has both supremum $\cup A_i$ and infimum $\cap A_i$.
\end{itemize}
In this case, for each pair $A, B \in \mathcal{A}$, we write $A \backslash B$ to denote $A \cap \neg B$; henceforth we often write $S \backslash A$ instead of $\neg A$. We refer to each $A \in \mathcal{A}$ as an {\it event}. In the special case that $\mathcal{A}$ is a collection of subsets of $S$ and $\supseteq$ is set containment, we say that $(S, \mathcal{A})$ is a {\it measurable space}.

\vspace{\baselineskip} The generalization from measurable spaces to $\sigma$-algebras is a minor technical detail that is not crucial to our main message, and we deliberately use set-theoretic notation to emphasize that this broader class is more familiar than exotic in the sense that many standard arguments involving set operations generalize (see Lemma~1 of \citealp{Mackenzie2019}). We relax the state space requirement not for the sake of generality, but in order to effectively deal with technical nuisances caused by null events in our proofs.\footnote{This is perhaps best illustrated by an example: if $(\mathcal{A}, \supseteq)$ is the collection of Lebesgue-measurable subsets of the state space $S = [0, 1]$ together with ordinary set containment, and if we declare two events to be equivalent if and only if the Lebesgue measure assigns zero to their symmetric difference, then (i)~both $(\mathcal{A}, \supseteq)$ and the associated {\it quotient} of equivalence classes $(\mathcal{A}', \supseteq')$ are $\sigma$-algebras, (ii)~we certainly want our results to cover the former, (iii)~it is convenient to derive results about the former by working with the latter because the latter has a unique null event, and (iv)~the latter is not isomorphic to any collection of subsets of a state space. We remark that this phenomenon is the reason that the representation theorem for $\sigma$-algebras (\citealp{Loomis1947}; \citealp{Sikorski1960}) has more nuance than the representation theorem for Boolean algebras \citep{Stone1936}.}

We use the term {\it assessment} to refer to a collection of event comparisons, and we focus on assessments that are (complete and transitive) event rankings.

\vspace{\baselineskip} \noindent \textsc{Definition:} Let $(\mathcal{A}, \supseteq)$ be a $\sigma$-algebra and let $\succsim$ be a binary relation on $\mathcal{A}$. We refer to $\succsim$ as an {\it assessment}, interpreting $A \succsim B$ to mean that $A$ is assessed to be at least as substantiated as $B$, and refer to $(\mathcal{A}, \supseteq, \succsim)$ as a {\it assessment space}. We say that $(\mathcal{A}, \supseteq, \succsim)$ satisfies
\begin{itemize}
\item {\it order} if and only if $\succsim$ is complete and transitive;

\item {\it separability} if and only if for each triple $A, B, C \in \mathcal{A}$ such that $A \cap C = B \cap C = \emptyset$, we have $A \succsim B$ if and only if $A \cup C \succsim B \cup C$;

\item {\it monotonicity} if and only if for each pair $A, B \in \mathcal{A}$ such that $A \subseteq B$, we have $B \succsim A$; and

\item {\it non-degeneracy} if and only if there are $A, B \in \mathcal{A}$ such that $A \succ B$.
\end{itemize}
We say that $(\mathcal{A}, \supseteq, \succsim)$ is an {\it inference} if and only if it satisfies the first two conditions, and a {\it qualitative probability} (\citealp{Bernstein1917}; \citealp{deFinetti1937}; \citealp{Koopman1940}) if and only if it satisfies all four.

\vspace{\baselineskip} As discussed in the introduction, qualitative probabilities fit the interpretation that events are assessed on the basis of relative likelihood, while inferences allow for the interpretation that events are assessed on the basis of how corroborated they are by a clue.

We also consider the following technical requirements.

\vspace{\baselineskip} \noindent \textsc{Definition:} An assessment space satisfies
\begin{itemize}
\item {\it continuity} if and only if for each $A \in \mathcal{A}$ and each $(B_i)_{i \in \mathbb{N}} \in \mathcal{A}^\mathbb{N}$ such that $B_1 \supseteq B_2 \supseteq ...$, (i)~if for each $	i \in \mathbb{N}$ we have $B_i \succsim A$, then $\cap B_i \succsim A$, and (ii)~if for each $i \in \mathbb{N}$ we have $A \succsim B_i$, then $A \succsim \cap B_i$;

\item {\it no atoms} if and only if for each $A \in \mathcal{A}$ such that $A \not \sim \emptyset$, there is $B \in \mathcal{A}$ such that $B \subseteq A$, $B \not \sim \emptyset$, and $B \not \sim A$; and

\item {\it absoluteness} if and only if $S \sim \emptyset$.
\end{itemize}
We say that an inference is moreover a {\it smooth inference} if and only if it satisfies {\it continuity} and {\it no atoms}.

\vspace{\baselineskip} The first axiom adapts the elegant {\it monotone continuity} of \cite{Villegas1964}, which notably was praised and adapted by \cite{Arrow1970}, to assessments that need not be monotonic. The second axiom, also due to \cite{Villegas1964}, is a richness requirement that rules out finite collections of events; these are known to be troublesome because they allow for qualitative probabilities without measure representations \citep{Kraft-Pratt-Seidenberg1959}. The last axiom is specifically written for inferences that are not qualitative probabilities: if the inference reflects a belief change from a prior to a posterior, then the axiom is implied so long as the largest event assessed in the inference is in fact the sure event.

\hypertarget{Section2.2}{}
\subsection{Representations}

We are interested in representing inferences, and our analysis involves the following classes of representations.

\vspace{\baselineskip} \noindent \textsc{Definition:} Abusing notation, let $\mathcal{A} = (\mathcal{A}, \supseteq)$ be a $\sigma$-algebra. We say that $\mu: \mathcal{A} \to \mathbb{R}$ satisfies
\begin{itemize}
\item {\it finite additivity} if and only if for each disjoint pair $A, B \in \mathcal{A}$, $\mu(A \cup B) = \mu(A) + \mu(B)$,

\item {\it countable additivity} if and only if for each sequence $(A_i)_{i \in \mathbb{N}} \in \mathcal{A}^\mathbb{N}$ such that $\{A_i\}_{i \in \mathbb{N}}$ is pairwise-disjoint, $\sum_{i \in \mathbb{N}} \mu(A_i)$ exists and is equal to $\mu(\cup A_i)$, and

\item {\it no measure-atoms} if and only if for each $A \in \mathcal{A}$ such that $\mu(A) \neq 0$, there is $B \in \mathcal{A}$ such that $B \subseteq A$ and $\mu(B) \not \in \{0, \mu(A)\}$, and

\item {\it probabilistic assignment} if and only if (i)~for each $A \in \mathcal{A}$ we have $\mu(A) \in [0, 1]$, and (ii)~$\mu(S) = 1$.
\end{itemize}
A function $\mu: \mathcal{A} \to \mathbb{R}$ that satisfies the fourth condition is a {\it probability measure} if and only if it satisfies the first condition, a {\it probability $\sigma$-measure} if and only if it satisfies the first two, and an {\it atomless probability $\sigma$-measure} if and only if it satisfies the first three; we denote these collections by $\mathbb{P}(\mathcal{A})$, $\mathbb{P}^\sigma(\mathcal{A})$, and $\mathbb{P}^\sigma_{\mathsf{na}}(\mathcal{A})$, respectively. Generalizing these notions, a function $\mu: \mathcal{A} \to \mathbb{R}$ is a {\it signed measure} if and only if it satisfies the first condition, a {\it signed $\sigma$-measure} if and only if it satisfies the first two, and an {\it atomless signed $\sigma$-measure} if and only if it satisfies the first three; we denote these collections by $\mathbb{S}(\mathcal{A})$, $\mathbb{S}^\sigma(\mathcal{A})$, and $\mathbb{S}^\sigma_{\mathsf{na}}(\mathcal{A})$, respectively. Finally, we let $\mu_{\mathsf{z}}$ denote the member of $\mathbb{S}^\sigma_{\mathsf{na}}(\mathcal{A})$ that maps each event to zero.

\vspace{\baselineskip} We remark that while some authors allow signed measures to assign $\infty$ or $-\infty$, our definition requires signed measures to always assign real numbers. We are interested in conditions that guarantee that an inference is compatible with a representation in the following sense.

\vspace{\baselineskip} \noindent \textsc{Definition:} Fix an assessment space and let $\mu: \mathcal{A} \to \mathbb{R}$ be a function. We say that $\mu$ is a {\it representation of $(\mathcal{A}, \supseteq, \succsim)$} if and only if for each pair $A, B \in \mathcal{A}$, we have $A \succsim B$ if and only if $\mu(A) \geq \mu(B)$.

\hypertarget{Section3}{}
\section{Results}

We begin with two classic results of Villegas. The first states that {\it continuity} is the ordinal analogue of countable additivity in the context of probability measures, while the second provides the foundation for representing beliefs with countably additive probabilities.

\hypertarget{TheoremV1}{}
\vspace{\baselineskip} \noindent \textsc{Theorem V1 \citep{Villegas1964}:}\footnote{This is Theorem~2 in Section~4 of \cite{Villegas1964}.} Fix a qualitative probability with representation $\mu \in \mathbb{P}(\mathcal{A})$. The qualitative probability satisfies {\it continuity} if and only if $\mu$ satisfies countable additivity.

\hypertarget{TheoremV2}{}
\vspace{\baselineskip} \noindent \textsc{Theorem V2 \citep{Villegas1964}:}\footnote{This is a restatement of Theorem~3 in Section~4 of \cite{Villegas1964} that follows directly from the original statement.} A qualitative probability satisfies {\it continuity} and {\it no atoms} if and only if it has a representation $\mu \in \mathbb{P}^\sigma_{\mathsf{na}}(\mathcal{A})$. In this case, there is no other representation in all of $\mathbb{P}(\mathcal{A}) \supseteq \mathbb{P}^\sigma_{\mathsf{na}}(\mathcal{A})$; $\mu$ is the unique probability measure representation.

\vspace{\baselineskip} Our first proposition verifies that we have adapted {\it continuity} properly, in the sense that it remains the ordinal analogue of countable additivity in the context of signed measures.

\vspace{\baselineskip} \noindent \textsc{Definition:} Fix an assessment space. We say that that a signed measure $\mu$ is {\it normalized} if and only if $\sup_{A \in \mathcal{A}} |\mu(A)| = \max_{A \in \mathcal{A}} |\mu(A)| = 1$. Observe that if an inference has a normalized signed measure representation, then it is non-degenerate.

\hypertarget{Proposition1}{}
\vspace{\baselineskip} \noindent \textsc{Proposition 1:} Fix an inference with normalized representation $\mu \in \mathbb{S}(\mathcal{A})$. The inference satisfies {\it continuity} if and only if $\mu$ satisfies countable additivity.

\vspace{\baselineskip} The proof is in \hyperlink{AppendixB}{Appendix~B}. Our first theorem generalizes \hyperlink{TheoremV2}{Theorem~V2} from qualitative probabilities to inferences by dropping both monotonicity.

\hypertarget{Theorem1}{}
\vspace{\baselineskip} \noindent \textsc{Theorem 1:} A {\it non-degenerate} inference is smooth if and only if it has a normalized representation $\mu \in \mathbb{S}^\sigma_{\mathsf{na}}(\mathcal{A})$. In this case, there is no other normalized representation in all of $\mathbb{S}(\mathcal{A}) \supseteq \mathbb{S}^\sigma_{\mathsf{na}}(\mathcal{A})$; $\mu$ is the unique normalized signed measure representation.

\vspace{\baselineskip} The proof is in \hyperlink{AppendixF}{Appendix~F}, and builds on results from  \hyperlink{AppendixA}{Appendix~A}, \hyperlink{AppendixC}{Appendix~C}, \hyperlink{AppendixD}{Appendix~D}, and \hyperlink{AppendixE}{Appendix~E}. See \hyperlink{Section4.4}{Section~4.4} for a proof sketch. Signed measures are well-studied objects in measure theory, and a well-known result about them immediately yields the result promised from the introduction that there is a top-ranked event.

\hypertarget{Corollary1}{}
\vspace{\baselineskip} \noindent \textsc{Corollary 1:} If an inference satisfies {\it continuity} and {\it no atoms}, then there is $A \in \mathcal{A}$ such that for each $B \in \mathcal{A}$, $A \succsim B$.

\vspace{\baselineskip} This follows from the fact that each signed measure has a Hahn decomposition \citep{Hahn1921}. In fact, we actually prove \hyperlink{Corollary1}{Corollary~1} (and more) before \hyperlink{Theorem1}{Theorem~1}. In particular, for the rest of the paper, we use {\it Hahn decomposition} to mean the following ordinal notion for inference that is analogous to the cardinal notion for signed measures.

\vspace{\baselineskip} \noindent \textsc{Definition:} Fix an assessment space. We say that $A \in \mathcal{A}$ is {\it non-negative} if $A \succsim \emptyset$ and {\it non-positive} if $\emptyset \succsim A$. A pair $(H^\oplus, H^\ominus) \in \mathcal{A} \times \mathcal{A}$ is a {\it Hahn decomposition} if and only if (i)~for each $A \in \mathcal{A}$, $H^\oplus \succsim A \succsim H^\ominus$, (ii)~$H^\oplus \cap H^\ominus = \emptyset$ and $H^\oplus \cup H^\ominus = S$, (iii)~each subevent of $H^\oplus$ is non-negative, and (iv)~each subevent of $H^\ominus$ is non-positive.

\vspace{\baselineskip} For intuition, it may be useful to imagine a Hahn decomposition as a separation of something that appears to have various shades of gray into a part that is pure black and a part that is pure white.\footnote{This suggestion is inspired by a quote of Rorschach from {\it Watchmen}: ``Black and white. Moving. Changing shape...but not mixing. No gray. Very, very beautiful."} In particular, consider again the measurable space from the illustrative example: $S$ is the unit circle and $\mathcal{A}$ is the collection of Lebesgue measurable subsets of $S$. Imagine the unit circle is partitioned into two events $H^\oplus$ and $H^\ominus$, with the former colored pure black and the latter colored pure white; this could be a simple design like a yin-yang or it could be extremely complicated. For each $A \in \mathcal{A}$, let $\mu(A)$ denote the difference between (i)~the Lebesgue measure of $A \cap H^\oplus$, and (ii)~the Lebesgue measure of $A \cap H^\ominus$. Then $\mu$ is a signed measure that captures an index of total darkness on the unit circle. Some such indices would be compatible with various shades of gray, but not $\mu$. Though this example is a bit too restrictive, because in general a Hahn decomposition is compatible with a third region of ``perfect gray" that neither contributes to darkness nor lightness, it should serve for intuition: to establish the existence of a Hahn decomposition, the technical problem is to separate the positive from the negative. In order to establish \hyperlink{Theorem1}{Theorem~1}, we first show that each smooth inference has a Hahn decomposition; see the proof sketch.

Our second theorem formalizes the intuition that an {\it absolute} smooth inference can be viewed as a change in beliefs with respect to a clue that is an assessed event. We do so using the notion of a Jordan decomposition of a signed measure \citep{Jordan1893} for the special case that the signed measure assigns zero to the largest event.

\vspace{\baselineskip} \noindent \textsc{Definition:} Fix an assessment space. For each $\mu \in \mathbb{S}^\sigma_{\mathsf{na}}(\mathcal{A})$ such that $\mu$ is normalized and $\mu(S)=0$, and for each pair $(\mu_0, \mu_1) \in \mathbb{P}^\sigma_{\mathsf{na}} \times \mathbb{P}^\sigma_{\mathsf{na}}$, we say that $(\mu_0, \mu_1)$ is a {\it Jordan decomposition of $\mu$} if and only if (i)~$\mu = \mu_1 - \mu_0$, and (ii)~there are disjoint supports $S_0, S_1 \in \mathcal{A}$ such that $\mu_0(S \backslash S_0) = \mu_1(S \backslash S_1) = 0$.

\vspace{\baselineskip} \noindent \textsc{Definition:} Fix an assessment space. We say that $(\mu_0, \mu_1, A^*) \in \mathbb{P}^\sigma_{\mathsf{na}} \times \mathbb{P}^\sigma_{\mathsf{na}} \in \mathcal{A}$ is a {\it Bayesian representation} if and only if (i)~for each pair $A, B \in \mathcal{A}$, $A \succsim B$ if and only if $\mu_1(A) - \mu_0(A) \geq \mu_1(B) - \mu_0(B)$, (ii)~$\mu_0(A^*) \in (0, 1)$, and (iii)~for each $A \in \mathcal{A}$, $\mu_1(A) = \frac{\mu_0(A \cap A^*)}{\mu_0(A^*)}$.

\hypertarget{Theorem2}{}
\vspace{\baselineskip} \noindent \textsc{Theorem 2:} A {\it non-degenerate} inference is smooth and {\it absolute} if and only if it has a normalized representation $\mu \in \mathbb{S}^\sigma_{\mathsf{na}}(\mathcal{A})$ such that $\mu(S) = 0$. In this case, $\mu$ is the unique normalized signed measure representation, there is a Hahn decomposition, and $\mu$ has a unique Jordan decomposition $(\mu^J_0, \mu^J_1) \in \mathbb{P}^\sigma_{\mathsf{na}} \times \mathbb{P}^\sigma_{\mathsf{na}}$. In particular, for each Hahn decomposition $(H^\oplus, H^\ominus)$ and each $A \in \mathcal{A}$,
\begin{align*}
\mu^J_0(A) &= -\inf \{\mu(B) | B \in \mathcal{A} \text{ and } B \subseteq A\} = -\mu(A \cap H^\ominus), \text{ and}
\\ \mu^J_1(A) &= \sup \{\mu(B) | B \in \mathcal{A} \text{ and } B \subseteq A\} = \mu(A \cap H^\oplus).
\end{align*}
Moreover, for each Hahn decomposition $(H^\oplus, H^\ominus)$, a tuple $(\mu_0, \mu_1, A^*) \in \mathbb{P}^\sigma_{\mathsf{na}} \times \mathbb{P}^\sigma_{\mathsf{na}} \times \mathcal{A}$ is a Bayesian representation if and only if
\begin{itemize}
\item $\mu_0 = (1 - \mu_0(H^\oplus)) \cdot \mu^J_0 + \mu_0(H^\oplus) \cdot \mu^J_1$,

\item $\mu_1 = \mu^J_1$,

\item $\mu_0(H^\oplus) \in (0, 1)$, and

\item $H^\oplus$ is equivalent to $A^*$ in the following senses: $A^* \sim H^\oplus$, $\mu_0(A^*) = \mu_0(H^\oplus)$, and $\mu_1(A^*) = \mu_1(H^\oplus)$.
\end{itemize}
Thus across all Bayesian representations, the posterior is unique, all clue guesses are in a suitable sense equivalent, and the prior is determined by the weight it assigns to each possible guess.

\vspace{\baselineskip} The proof, which uses \hyperlink{Theorem1}{Theorem~1}, is in \hyperlink{AppendixG}{Appendix~G}. Finally, it would be poor form to neglect the Radon-Nikodym theorem (\citealp{Radon1913}; \citealp{Nikodym1930}) in a discussion of signed measures, Hahn decompositions, and Jordan decompositions.

\hypertarget{Corollary2}{}
\vspace{\baselineskip} \noindent \textsc{Corollary 2:} For each {\it non-degenerate} inference that is smooth and {\it absolute} such that $(S, \mathcal{A})$ is a measurable space, and for each Bayesian representation $(\mu_0, \mu_1, A^*)$, there is a density function $f:S \to \mathbb{R}$ such that for each $A \in \mathcal{A}$, $\mu_1(A) = \int_A f d\mu_0$.

\vspace{\baselineskip} \noindent \textsc{Proof:} Assume the hypotheses. By \hyperlink{Theorem2}{Theorem~2}, for each $A \in \mathcal{A}$ such that $\mu_0(A) = 0$, we have $\mu_1(A) = 0$, so the conclusion follows from the Radon-Nikodym theorem.~$\blacksquare$

\hypertarget{Section4}{}
\section{Technique}

The rest of the paper is dedicated to technique. We begin by introducing a few key notions that are pervasive throughout our arguments.

\hypertarget{Section4.1}{}
\subsection{Sign language}

Because we drop monotonicity, we allow for negative events, and this in turn introduces some novel nuance about positive events and null events.

\vspace{\baselineskip} \noindent \textsc{Definition:} Fix an inference. First, we say that $A \in \mathcal{A}$ is (i)~{\it positive} if $A \succ \emptyset$, (ii)~{\it non-negative} if $A \succsim \emptyset$, (iii)~{\it null} if and only if $A \sim \emptyset$, (iv)~{\it non-positive} if and only if $\emptyset \succsim A$, and {\it negative} if and only if $\emptyset \succ A$. Second, we say that an event is (i)~{\it purely non-negative} if each of its subevents (including itself) is non-negative, (ii)~{\it purely null} if each of its subevents (including itself) is null, and (iii)~{\it purely non-positive} if each of its subevents (including itself) is non-positive. Finally, we say that an event is (i)~{\it thoroughly positive} if and only if it is both positive and purely non-negative, and (ii)~{\it thoroughly negative} if and only if it is both negative and purely non-positive.

\vspace{\baselineskip} Similarly, there is now sign-related nuance for collections of events.

\vspace{\baselineskip} \noindent \textsc{Definition:} Fix an inference. We say that $\mathcal{A}' \subseteq \mathcal{A}$ is (i)~{\it purely positive} if and only if it includes only positive events, (ii)~{\it positive} if and only if it includes some positive event and no negative event, (iii)~{\it negative} if and only if it includes some negative event and no positive event, and (iv)~{\it purely negative} if and only if it includes only negative events.

\vspace{\baselineskip} Finally, there is now sign-related nuance for sequences of events.

\vspace{\baselineskip} \noindent \textsc{Definition:} Fix an inference. We say that $(A_i) \in \mathcal{A}^\mathbb{N}$ is (i)~{\it shrinking} if and only if $A_1 \supseteq A_2 \supseteq ...$, (ii)~{\it growing} if and only if $A_1 \subseteq A_2 \subseteq ...$, and (iii)~{\it vanishing} if and only if it is shrinking and $\cap A_i = \emptyset$. We say that a vanishing sequence is (i)~{\it descending} if and only if $A_1 \succ A_2 \succ ... \succ \emptyset$, and (ii)~{\it ascending} if and only if $A_1 \prec A_2 \prec ... \prec \emptyset$.

\hypertarget{Section4.2}{}
\subsection{Knives}

In much of the literature on fairly dividing a cake \citep{Steinhaus1948}, there is a knife that glides continuously over the cake that can stop and cut at any time, and this is extremely useful for construction. We intend to suggest such an object with the following definition.

\vspace{\baselineskip} \noindent \textsc{Definition:} Fix a $\sigma$-algebra. For each function $\mu: \mathcal{A} \to \mathbb{R}$ and each event $A \in \mathcal{A}$, an {\it $(A|\mu)$-knife} is an indexed list of events $(\kappa_v)_{v \in [0, \mu(A)]} \in \mathcal{A}^{[0, \mu(A)]}$ such that (i)~for each pair $v, v' \in [0, \mu(A)]$ such that $v' > v$, we have $A \supseteq \kappa_{v'} \supseteq \kappa_v$, and (ii)~for each $v \in [0, \mu(A)]$, we have $\mu(\kappa_v) = v$.

\hypertarget{TheoremS}{}
\vspace{\baselineskip} \noindent \textsc{Theorem S \citep{Sierpinski1922}:}\footnote{This well-known result is generally attributed to \cite{Sierpinski1922}, which includes a similar result.} Fix a $\sigma$-algebra and let $\mu \in \mathbb{P}^\sigma_{\mathsf{na}}(\mathcal{A})$. For each $A \in \mathcal{A}$, there is an $(A|\mu)$-knife.

\hypertarget{Section4.3}{}
\subsection{Event collections}

For the purposes of construction, it is sometimes useful to partition an event into many pieces and then assemble those pieces.

\vspace{\baselineskip} \noindent \textsc{Definition:} Fix an assessment space. We say that $\mathcal{A}' \subseteq \mathcal{A}$ is {\it pairwise-disjoint}, or equivalently say that $\mathcal{A}'$ is an {\it antichain}, if and only if for each distinct pair $A, B \in \mathcal{A}$, $A \cap B = \emptyset$. We say that $\mathcal{M} \subseteq \mathcal{A}$ is a {\it mosaic} if and only if it is an antichain of non-null events,\footnote{To avoid confusion: the term {\it mosaic} is also used for a generalization of Boolean algebra by \cite{Kopylov2007}; the notions are unrelated. We recycle the term to suggest a disjoint tiling by pieces of substance.} and say that it is a {\it $\sigma$-mosaic} if and only if it is a mosaic and $|\mathcal{M}| = |\mathbb{N}|$.

\vspace{\baselineskip} Before proceeding, we remark that there are several closely related notions. When defining the countable chain condition, \cite{Halmos1963} implicitly requires an antichain to moreover be a collection of nonempty events; in this case the usual notion of partition is a maximal antichain. Deviating from that usual notion, \cite{Villegas1964} defines an incomplete partition to be any collection of non-null events such that each distinct pair has a null intersection, then defines a partition to be a maximal incomplete partition. Our definition of antichain matches that of \cite{Balcar-Jech2006}, and we selected these notions instead of alternatives based on the needs of our proof.

Finally, we consider the usual collection of events that are informed by the agent's comparisons.

\vspace{\baselineskip} \noindent \textsc{Definition:} Fix an inference. For each $A \in \mathcal{A}$, we define the {\it upper contour set of $A$} by $\mathcal{UCS}(A) \equiv \{B \in \mathcal{A} | B \succsim A\}$ and we define the {\it lower contour set of $A$} by $\mathcal{LCS}(A) \equiv \{B \in \mathcal{A} | A \succsim B\}$.

\hypertarget{Section4.5}{}
\subsection{Proof sketch}

The proofs span seven appendices. After establishing some basic lemmas that are freely used throughout the rest of the paper (\hyperlink{AppendixA}{Appendix~A}) and establishing our earlier claim that we have properly adapted continuity to non-monotonic assessments (\hyperlink{AppendixB}{Appendix~B}), we dedicate three appendices to the existence of a Hahn decomposition. Assume throughout this proof sketch that unless otherwise specified, we are assuming the inference is both non-degenerate and smooth, though for some of these results less is assumed.

We begin by gathering some basic lemmas about {\it continuity}, antichains, and mosaics (\hyperlink{AppendixC}{Appendix~C}), and already, we encounter challenges following the path charted by \cite{Villegas1964}. For example, under monotonicity, {\it no atoms} allows us to split a positive event $A$ into two smaller parts that are between $\emptyset$ and $A$, so iterative application of {\it no atoms} allows us to construct a vanishing sequence with arbitrarily small positive events. Without monotonicity, however, {\it no atoms} only allows us to split a positive event $A$ into two parts that are not equivalent to $A$ or $\emptyset$, and it may be that one is above $A$ and one is below $\emptyset$. Though we would like both a descending vanishing sequence with arbitrarily small positive events and an ascending vanishing sequence with arbitrarily small negative events, at this stage we are only able to guarantee one or the other (\hyperlink{Lemma7}{Lemma~7} and \hyperlink{Lemma8}{Lemma~8}), and we use this to establish that in an interesting case with both a positive event and a negative event, there is an event that is either thoroughly positive or thoroughly negative (\hyperlink{Proposition2}{Proposition~2}). Loosely, if without loss of generality we suppose that there is a descending vanishing sequence, then the proof of \hyperlink{Proposition2}{Proposition~2} involves using arbitrary small positive events to iteratively remove ``big enough" positive regions from a negative event, with increasingly flexible standards for ``big enough," until something yet remains but there is nothing positive left to remove. In the yin-yang metaphor, this is like iteratively removing ink from a region that is not pure black, by cutting off pieces, until what remains is an inkless piece. Of course, {\it continuity} is crucial here.

Next, we establish \hyperlink{Proposition3}{Proposition~3}, which states that under the additional assumption that there is a unique purely null event, there is a Hahn decomposition (\hyperlink{AppendixD}{Appendix~D}). First, after ruling out easy cases, by \hyperlink{Proposition2}{Proposition~2} there is either a thoroughly positive event or a thoroughly negative event, and without loss of generality there is a thoroughly positive event. Second, since there is a unique null event, we can show that the $\sigma$-algebra satisfies the countable chain condition. This implies that the $\sigma$-algebra is a complete lattice, but we actually use something slightly stronger from a proof of \cite{Halmos1963}: each collection of events $\mathcal{A}' \subseteq \mathcal{A}$ has a countable subcollection $\mathcal{A}'' \subseteq \mathcal{A}'$ such that $\mathcal{A}'$ and $\mathcal{A}''$ have the same upper bounds. We apply this to the nonempty collection of thoroughly positive events to construct the Hahn decomposition.

To finally show that there is a Hahn decomposition in general (\hyperlink{Proposition4}{Proposition~4}), we show that the additional assumption that there is a unique purely null event can be relaxed (\hyperlink{AppendixE}{Appendix~E}). The basic technique, which was also used in \cite{Mackenzie2019}, is to take the quotient of the original assessment space with respect to a $\sigma$-ideal of events, then argue that the latter inherits properties from the former, and to do so such that the latter has a unique null event. Then by \hyperlink{Proposition3}{Proposition3}, the quotient space has a Hahn decomposition, so the original space does. Again, the lack of monotonicity creates nuance. The collection of null events is no longer a $\sigma$-ideal, because a subevent of a null event need not be null, but the technique can be salvaged by instead working with the collection of purely null events.

The existence of a Hahn decomposition, in turn, allows us to establish \hyperlink{Theorem1}{Theorem~1}, which promises a signed measure representation (\hyperlink{AppendixF}{Appendix~F}). In the interesting case where there is both a positive event and a negative event, the basic idea is to apply \hyperlink{TheoremV2}{Theorem~V2} to both (i)~the subevents of $H^\oplus$ with the given ranking, and (ii)~the subevents of $H^\ominus$ with the {\it reverse} of the given ranking, in order to obtain probability measures $\mu^\oplus$ and $\mu^\ominus$. In order to calibrate them, we use \hyperlink{TheoremS}{Theorem~S} to associate the ``smaller" Hahn component with an {\it annulment} in the ``larger" Hahn component, by which we mean that the union of the two is null. This yields a scale with which to combine $\mu^\oplus$ and $\mu^\ominus$ into a candidate representation $\mu$, and a key step involves verifying if an event contained in $H^\oplus$ and an event contained in $H^\ominus$ have opposite measures, then their union is null.

Finally, we establish \hyperlink{Theorem2}{Theorem~2}, which states that under {\it absoluteness} we have Bayesian representations that are closely related to the unique Jordan decomposition $(\mu^J_0, \mu^J_1)$ of the unique signed measure representation (\hyperlink{AppendixG}{Appendix~G}). The bulk of the proof involves relating an arbitrary Bayesian representation $(\mu_0, \mu_1, A^*)$ to $(\mu^J_0, \mu^J_1)$ and an arbitrary Hahn decomposition $(H^\oplus, H^\ominus)$. At a high level, the idea is to relate $A$ and $H^\oplus$, then use this relationship to in turn relate six measures: $\mu_0$, $\mu_1$, $\mu_1 - \mu_0$, $\mu^J_0$, $\mu^J_1$, and $\mu^J_1 - \mu^J_0$.

\appendix
\setcounter{secnumdepth}{0}
\hypertarget{AppendixA}{}
\section{Appendix A - Basic lemmas}

In this appendix, we prove three basic lemmas (\hyperlink{Lemma1}{Lemma~1}, \hyperlink{Lemma2}{Lemma~2}, and \hyperlink{Lemma3}{Lemma~3}). For brevity, to avoid repeated reference to these basic lemmas throughout the rest of the proof, after this appendix we simply say ``by {\it separability}" when invoking the first two lemmas and ``by {\it continuity}" when invoking to the third.

\hypertarget{Lemma1}{}
\vspace{\baselineskip} \noindent \textsc{Lemma 1:} Fix an inference. For each four $A, B, A', B' \in \mathcal{A}$, if (i)~$A \succsim B$, (ii)~$A' \succsim B'$, (iii)~$A \cap A' = \emptyset$, and (iv)~$B \cap B' = \emptyset$, then $A \cup A' \succsim B \cup B'$. If moreover $A \succ B$, then $A \cup A' \succ B \cup B'$.

\vspace{\baselineskip} This lemma is a variant of Exercise 5a of \cite{Savage1972}. The latter imposes the stronger hypothesis that $\succsim$ is a qualitative probability but does not impose our hypothesis that $B \cap B' = \emptyset$, and if we omit both hypotheses, the statement is false.\footnote{Let $S = \{s_1, s_2, s_3, s_4, s_5\}$ and $\mathcal{A} = 2^S$. Define $u: S \to \mathbb{R}$ by $u(s_1) = 1$, $u(s_2) = 1$, $u(s_3) = 3$, $u(s_4) = 3$, and $u(s_5) = -3$. Let $\succsim$ be such that for each pair $A, B \in \mathcal{A}$, we have $A \succsim B$ if and only if $\sum_{s \in A} u(s) \geq \sum_{s \in B} u(s)$. Define $A \equiv \{s_1\}$, $A' \equiv \{s_2\}$, $B \equiv \{s_3, s_5\}$, and $B' \equiv \{s_4, s_5\}$. Then we have $A \succsim B$, $A' \succsim B'$, and $A \cap A' = \emptyset$, but $B \cup B' \succ A \cup A'$.} That said, the proof outlined in Savage's hint for the exercise suffices for our lemma, and we include it to confirm that the conclusion does not require {\it monotonicity}.

\vspace{\baselineskip} \noindent \textsc{Proof:} Let $A$, $B$, $A'$, and $B'$ satisfy the hypotheses of the first statement. Using {\it separability} twice, $(A \backslash B') \cup A' \succsim (A \backslash B') \cup B' = A \cup (B' \backslash A) \succsim B \cup (B' \backslash A)$, so by {\it separability} we have $A \cup A' = ((A \backslash B') \cup A') \cup (A \cap B') \succsim (B \cup (B' \backslash A)) \cup (A \cap B') = B \cup B'$, as desired. If moreover $A \succ B$, then we obtain the desired conclusion by replacing the second and third instance of $\succsim$ with $\succ$ in the preceding sentence.~$\blacksquare$

\hypertarget{Lemma2}{}
\vspace{\baselineskip} \noindent \textsc{Lemma 2:} Fix an inference. For each triple $A, B, B' \in \mathcal{A}$ such that $B \subseteq A$ and $B' \subseteq A$, we have $B \succsim B'$ if and only if $A \backslash B' \succsim A \backslash B$.

\vspace{\baselineskip} This is a slight extension of Exercise~3 of \cite{Savage1972}. Again, we include the proof simply to confirm that the conclusion does not require {\it monotonicity}.

\vspace{\baselineskip} \noindent \textsc{Proof:} Let $A$, $B$, and $B'$ satisfy the hypotheses. We cannot have $B \succsim B'$ and $A \backslash B \succ A \backslash B'$; else by \hyperlink{Lemma1}{Lemma~1} we have $A = B \cup (A \backslash B) \succ B' \cup (A \backslash B') = A$, contradicting $A \sim A$. By a similar argument, we cannot have $B' \succ B$ and $A \backslash B' \succsim A \backslash B$.~$\blacksquare$

\hypertarget{Lemma3}{}
\vspace{\baselineskip} \noindent \textsc{Lemma 3:} Fix a {\it continuous} inference. For each $A \in \mathcal{A}$ and each $(B_i)_{i \in \mathbb{N}} \in \mathcal{A}^\mathbb{N}$ such that $B_1 \subseteq B_2 \subseteq ...$, (i)~if for each $	i \in \mathbb{N}$ we have $B_i \succsim A$, then $\cup B_i \succsim A$, and (ii)~if for each $i \in \mathbb{N}$ we have $A \succsim B_i$, then $A \succsim \cup B_i$.

\vspace{\baselineskip} \noindent \textsc{Proof:} Let $A$ and $(B_i)_{i \in \mathbb{N}}$ satisfy the hypotheses.

First, assume that for each $i \in \mathbb{N}$ we have $B_i \succsim A$. By \hyperlink{Lemma2}{Lemma~2}, for each $i \in \mathbb{N}$ we have $S \backslash A \succsim S \backslash B_i$, so $(S \backslash B_i)_{i \in \mathbb{N}}$ is a shrinking sequence of events $\mathcal{LCS}(S \backslash A)$, so by {\it continuity} we have $S \backslash A \succsim \cap (S \backslash B_i) = S \backslash (\cup B_i)$, so by \hyperlink{Lemma2}{Lemma~2} we have $\cup B_i \succsim A$.

Second, assume that for each $i \in \mathbb{N}$ we have $A \succsim B_i$. By \hyperlink{Lemma2}{Lemma~2}, for each $i \in \mathbb{N}$ we have $S \backslash B_i \succsim S \backslash A$, so $(S \backslash B_i)_{i \in \mathbb{N}}$ is a shrinking sequence of events in $\mathcal{UCS}(S \backslash A)$, so by {\it continuity} we have $S \backslash (\cup B_i) = \cap (S \backslash B_i) \succsim S \backslash A$, so by \hyperlink{Lemma2}{Lemma~2} we have $A \succsim \cup B_i$.~$\blacksquare$

\hypertarget{AppendixB}{}
\section{Appendix B - Continuity and countable additivity}

In this appendix, we prove \hyperlink{Proposition1}{Proposition~1}.

\vspace{\baselineskip} \noindent \textsc{Proposition 1:} Fix an inference with normalized representation $\mu \in \mathbb{S}(\mathcal{A})$. The inference satisfies {\it continuity} if and only if $\mu$ satisfies countable additivity.

\vspace{\baselineskip} \noindent \textsc{Proof:} It is straightforward to show that if $\mu$ is countably additive, then the inference is {\it continuous}; we omit the argument. Thus let us assume the inference is {\it continuous}. At a high level, the proof approach is to organize the argument into old cases handled by the proof of \cite{Villegas1964} and new cases that are simpler than the old cases.

By the proof of 326K in Chapter~32 of \cite{Fremlin2012},\footnote{Fremlin's statement uses non-increasing sequences and ours uses weakly decreasing sequences, which are distinct because a $\sigma$-algebra is a partial order, but Fremlin's short proof applies to both statements.} $\mu$ is countably additive if and only if for each vanishing $(V_i) \in \mathcal{A}^\mathbb{N}$ we have $\lim \mu(V_i) = 0$. To establish the latter, let $(V_i) \in \mathcal{A}^\mathbb{N}$ be vanishing and let $\varepsilon > 0$. Let $\mathcal{A}' \equiv \{A_i\}$ denote the antichain $\{V_i \backslash V_{i+1}\}_{i \in \mathbb{N}}$, and define (i)~$\mathcal{A}^+ \equiv \{A \in \mathcal{A}' | A \succ \emptyset\}$, and (ii)~$\mathcal{A}^- \equiv \{A \in \mathcal{A}' | A \prec \emptyset\}$.

First, we claim that there is $i \in \mathbb{N}$ such that for each $j \geq i$, $\varepsilon > \mu(V_j)$. Indeed, if $\mathcal{A}^+$ is finite, then there is $i \in \mathbb{N}$ such that for each $j \geq i$, $V_j \precsim V_{j+1} \precsim...$, so by {\it continuity} we have $\emptyset = \cap V_k \succsim V_j$, so $\varepsilon > 0 = \mu(\emptyset) \geq \mu(V_j)$, as desired; thus let us assume that $\mathcal{A}^+$ is infinite. In this case, since $\mu$ is normalized, there is $A^+ \in \mathcal{A}^+$ such that $\mu(A^+) \in (0, \varepsilon)$. There cannot be an infinite index set $I \subseteq \mathbb{N}$ such that for each $i \in I$ we have $V_i \succsim A^+$; else by {\it continuity} we would have $\emptyset = \cap_{i \in I} V_i \succsim A^+ \succ \emptyset$, contradicting $\emptyset \sim \emptyset$. Thus there is $i \in \mathbb{N}$ such that for each $j \geq i$, we have $A^+ \succ V_j$ and thus $\varepsilon > \mu(A^+) > \mu(V_j)$, as desired.

Second, we claim that there is $i \in \mathbb{N}$ such that for each $j \geq i$, $-\varepsilon < \mu(V_j)$. Indeed, if $\mathcal{A}^-$ is finite, then there is $i \in \mathbb{N}$ such that for each $j \geq i$, $V_j \succsim V_{j+1} \succsim...$, so by {\it continuity} we have $\emptyset = \cap V_k \precsim V_j$, so $-\varepsilon < 0 = \mu(\emptyset) \leq \mu(V_j)$, as desired; thus let us assume that $\mathcal{A}^-$ is infinite. In this case, since $\mu$ is normalized, there is $A^- \in \mathcal{A}^-$ such that $\mu(A^-) \in (-\varepsilon, 0)$. There cannot be an infinite index set $I \subseteq \mathbb{N}$ such that for each $i \in I$ we have $V_i \precsim A^-$; else by {\it continuity} we would have $\emptyset = \cap_{i \in I} V_i \precsim A^- \prec \emptyset$, contradicting $\emptyset \sim \emptyset$. Thus there is $i \in \mathbb{N}$ such that for each $j \geq i$, we have $A^- \prec V_j$ and thus $-\varepsilon < \mu(A^-) < \mu(V_j)$, as desired.

To conclude, by the two claims, there is $i \in \mathbb{N}$ such that for each $j \geq i$, $\mu(V_j) \in (-\varepsilon, \varepsilon)$; thus $\lim \mu(V_i) = 0$, as desired.~$\blacksquare$

\hypertarget{AppendixC}{}
\section{Appendix C - Mosaics and thoroughly signed events}

In this appendix, we prove \hyperlink{Proposition2}{Proposition~2}. To do so, we first prove five lemmas that collectively concern continuity, antichains, and mosaics (\hyperlink{Lemma4}{Lemma~4}, \hyperlink{Lemma5}{Lemma~5}, \hyperlink{Lemma6}{Lemma~6}, \hyperlink{Lemma7}{Lemma~7}, and \hyperlink{Lemma8}{Lemma~8}).

\hypertarget{Lemma4}{}
\vspace{\baselineskip} \noindent \textsc{Lemma 4:} Fix a {\it continuous} inference. For each antichain $\mathcal{A}' \subseteq \mathcal{A}$, each $A^+ \in \mathcal{A}$ such that $A^+ \succ \emptyset$, and each $A^- \in \mathcal{A}$ such that $\emptyset \succ A^-$, both $\mathcal{A}' \cap \mathcal{UCS}(A^+)$ and $\mathcal{A}' \cap \mathcal{LCS}(A^-)$ are finite.

\vspace{\baselineskip} \noindent \textsc{Proof:} Let $\mathcal{A}'$, $A^+$, and $A^-$ satisfy the hypotheses.

First, assume by way of contradiction that $\mathcal{A}' \cap \mathcal{UCS}(A^+)$ is infinite. Then there is $\{A_i\}_{i \in \mathbb{N}} \subseteq \mathcal{A}' \cap \mathcal{UCS}(A^+)$, which is pairwise-disjoint as a subset of an antichain. For each pair $a, b \in \mathbb{N}$ such that $b>a$, define $B_{[a, b]} \equiv \cup_{i \in \{a, a+1, ..., b\}} A_i$; by iterative application of {\it separability} we have $B_{[a, b]} \succsim A^+$ and thus $B_{[a, b]} \in \mathcal{UCS}(A^+)$. For each $a \in \mathbb{N}$, define $C_a \equiv \cup_{b \geq a} B_{[a, b]}$; this is the supremum of a growing sequence in $\mathcal{UCS}(A^+)$, so by {\it continuity} we have $C_a \in \mathcal{UCS}(A^+)$. But then $(C_a)_{a \in \mathbb{N}}$ is a shrinking sequence in $\mathcal{UCS}(A^+)$, so by {\it continuity} we have $\emptyset = \cap C_a \in \mathcal{UCS}(A^+)$ and thus $\emptyset \succsim A^+$, contradicting $A^+ \succ \emptyset$.

Second, assume by way of contradiction that $\mathcal{A}' \cap \mathcal{LCS}(A^-)$ is infinite. Then there is $\{A_i\}_{i \in \mathbb{N}} \subseteq \mathcal{A}' \cap \mathcal{LCS}(A^-)$, which is pairwise-disjoint as a subset of an antichain. For each pair $a, b \in \mathbb{N}$ such that $b>a$, define $B_{[a, b]} \equiv \cup_{i \in \{a, a+1, ..., b\}} A_i$; by iterative application of {\it separability} we have $A^- \succsim B_{[a, b]}$ and thus $B_{[a, b]} \in \mathcal{LCS}(A^-)$. For each $a \in \mathbb{N}$, define $C_a \equiv \cup_{b \geq a} B_{[a, b]}$; this is the supremum of a growing sequence in $\mathcal{LCS}(A^-)$, so by {\it continuity} we have $C_a \in \mathcal{LCS}(A^-)$. But then $(C_a)_{a \in \mathbb{N}}$ is a shrinking sequence in $\mathcal{LCS}(A^-)$, so by {\it continuity} we have $\emptyset = \cap C_a \in \mathcal{LCS}(A^-)$ and thus $A^- \succsim \emptyset$, contradicting $\emptyset \succ A^-$.~$\blacksquare$

\hypertarget{Lemma5}{}
\vspace{\baselineskip} \noindent \textsc{Lemma 5:} Fix a {\it continuous} inference. Each mosaic is countable.

\vspace{\baselineskip} \noindent \textsc{Proof:} Let $\mathcal{M}$ be a mosaic, let $\mathcal{M}^+ \subseteq \mathcal{M}$ be the collection of positive events in $\mathcal{M}$, and let $\mathcal{M}^- \subseteq \mathcal{M}$ be the collection of negative events in $\mathcal{M}$. Since $\mathcal{M}$ is a mosaic, thus $\mathcal{M} = \mathcal{M}^+ \cup \mathcal{M}^-$. By \hyperlink{Lemma4}{Lemma~4}, for each $A \in \mathcal{M}^+$ we have that $\mathcal{M}^+ \cap \mathcal{UCS}(A)$ is finite, so $\mathcal{M}^+ = \cup_{A \in \mathcal{M}^+} (\mathcal{M}^+ \cap \mathcal{UCS}(A))$ is a countable union of finite sets and thus countable. Similarly, by \hyperlink{Lemma4}{Lemma~4}, for each $A \in \mathcal{M}^-$ we have that $\mathcal{M}^- \cap \mathcal{LCS}(A)$ is finite, so $\mathcal{M}^- = \cup_{A \in \mathcal{M}^-} (\mathcal{M}^- \cap \mathcal{LCS}(A))$ is a countable union of finite sets and thus countable. Altogether, then, $\mathcal{M} = \mathcal{M}^+ \cup \mathcal{M}^-$ is countable as a finite union of countable sets, as desired.~$\blacksquare$

\hypertarget{Lemma6}{}
\vspace{\baselineskip} \noindent \textsc{Lemma 6:} Fix a {\it continuous} inference and let $\mathcal{A}' \subseteq \mathcal{A}$ be a countable antichain. If each event in $\mathcal{A}'$ is non-negative, then $\cup \mathcal{A}'$ is non-negative, and if moreover some event in $\mathcal{A}'$ is positive, then $\cup \mathcal{A}'$ is positive. Similarly, if each event in $\mathcal{A}'$ is non-positive, then $\cup \mathcal{A}'$ is non-positive, and if moreover some event in $\mathcal{A}'$ is negative, then $\cup \mathcal{A}'$ is negative.

\vspace{\baselineskip} \noindent \textsc{Proof:} Let $\mathcal{A}'$ be a countable antichain. Since $\mathcal{A}'$ is countable, thus we can index it $\mathcal{A}' = \{A_i\}_{i \in I}$ using index set $I \subseteq \mathbb{N}$ such that if $\mathcal{A}'$ includes a non-null event, then $1 \in I$ and $A_1$ is non-null. For each $i \in \mathbb{N}$, define $B_i \equiv \cup_{j \leq i} A_j$.

If each event in $\mathcal{A}'$ is non-negative, then by iterative application of {\it separability} we have that for each $i \in \mathbb{N}$, $B_i \succsim \emptyset$; thus by {\it continuity} we have that $\cup \mathcal{A}' = \cup B_i \succsim \emptyset$, so $\cup \mathcal{A}'$ is non-negative. In this case, if moreover some event in $\mathcal{A}'$ is positive, then $A_1$ is positive, and by iterative application of {\it separability} we have that for each $i \in \mathbb{N}$, $B_i \succsim A_1$; thus by {\it continuity} we have that $\cup \mathcal{A}' = \cup B_i \succsim A_1 \succ \emptyset$, so $\cup \mathcal{A}'$ is positive.

If each event in $\mathcal{A}'$ is non-positive, then by iterative application of {\it separability} we have that for each $i \in \mathbb{N}$, $B_i \precsim \emptyset$; thus by {\it continuity} we have that $\cup \mathcal{A}' = \cup B_i \precsim \emptyset$, so $\cup \mathcal{A}'$ is non-positive. In this case, if moreover some event in $\mathcal{A}'$ is negative, then $A_1$ is negative, and by iterative application of {\it separability} we have that for each $i \in \mathbb{N}$, $B_i \precsim A_1$; thus by {\it continuity} we have that $\cup \mathcal{A}' = \cup B_i \precsim A_1 \prec \emptyset$, so $\cup \mathcal{A}'$ is negative.~$\blacksquare$

\hypertarget{Lemma7}{}
\vspace{\baselineskip} \noindent \textsc{Lemma 7:} Fix a smooth inference. For each $A \in \mathcal{A}$ such that $A \not \sim \emptyset$, there is a $\sigma$-mosaic $\mathcal{M}$ that (i)~consists of subevents of $A$, and (ii)~is either purely positive or purely negative.

\vspace{\baselineskip} \noindent \textsc{Proof:} Let $A$ satisfy the hypothesis and define $A_1 \equiv A$; then $A_1 \not \sim \emptyset$. For each $i \in \mathbb{N}$ such that $A_i \not \sim \emptyset$, we choose $B_i \in \mathcal{A}$ and $A_{i+1} \in \mathcal{A}$ as follows. First, by {\it no atoms}, there is $B_i \in \mathcal{A}$ such that $B_i \subseteq A_i$, $B_i \not \sim \emptyset$, and $B_i \not \sim A_i$. Define $A_{i+1} \equiv A_i \backslash B_i$. We cannot have $A_{i+1} \sim \emptyset$; else by {\it separability} $A_i = B_i \cup A_{i+1} \sim B_i$, contradicting $B_i \not \sim A_i$.

Define $\mathcal{M}^* \equiv \{B_i\}_{i \in \mathbb{N}}$. By construction, $\mathcal{M}^*$ is a $\sigma$-mosaic of subevents of $A$. Let $\mathcal{M}^+$ be the collection of positive members of $\mathcal{M}^*$ and let $\mathcal{M}^-$ be the collection of negative members of $\mathcal{M}^*$; at least one of these is infinite and can be selected as the desired $\mathcal{M}$.~$\blacksquare$

\hypertarget{Lemma8}{}
\vspace{\baselineskip} \noindent \textsc{Lemma 8:} Fix a smooth inference. For each purely positive $\sigma$-mosaic $\mathcal{M}$, there is a descending vanishing sequence whose first event is $\cup \mathcal{M}$. For each purely negative $\sigma$-mosaic $\mathcal{M}$, there is an ascending vanishing sequence whose first event is $\cup \mathcal{M}$.

\vspace{\baselineskip} \noindent \textsc{Proof:} Let $\mathcal{M}$ satisfy the hypothesis and let $\{A_i\}_{i \in \mathbb{N}}$ be an arbitrary indexing of $\mathcal{M}$. For each $i \in \mathbb{N}$, define $V_i \equiv \cup_{j \geq i} A_i$. It is easy to verify that $\cup \mathcal{M} = V_1 \supseteq V_2 \supseteq ...$ and $\cap V_i = \emptyset$. If $\mathcal{M}$ is purely positive, then for each $i \in \mathbb{N}$ we have that $\{A_j\}_{j \geq i}$ is a countable antichain of positive events, so by \hyperlink{Lemma6}{Lemma~6} we have $V_i \succ \emptyset$; thus by {\it separability}, for each $i \in \mathbb{N}$ we have $V_i \succ V_{i+1} \succ \emptyset$, so $(V_i)$ is descending. Similarly, if $\mathcal{M}$ is purely negative, then for each $i \in \mathbb{N}$ we have that $\{A_j\}_{j \geq i}$ is a countable antichain of negative events, so by \hyperlink{Lemma6}{Lemma~6} we have $V_i \prec \emptyset$; thus by {\it separability}, for each $i \in \mathbb{N}$ we have $V_i \prec V_{i+1} \prec \emptyset$, so $(V_i)$ is ascending.~$\blacksquare$

\hypertarget{Proposition2}{}
\vspace{\baselineskip} \noindent \textsc{Proposition 2:} Fix a smooth inference. At least one of the following is true: (i)~each positive event contains a thoroughly positive event, or (ii)~each negative event contains a thoroughly negative event.

\vspace{\baselineskip} \noindent \textsc{Proof:} If there is no positive event, or if there is no negative event, then the conclusion is immediate; thus let us assume that there are a positive event and a negative event. By \hyperlink{Lemma7}{Lemma~7} and \hyperlink{Lemma8}{Lemma~8}, there is a vanishing sequence $(V_i)$ that is either descending or ascending.

Observe that if we establish the lemma for the case that $(V_i)$ is descending, then we can easily prove the lemma for the case that $(V_i)$ is ascending as follows: (i)~suppose that $(V_i)$ is ascending, (ii)~define $\succsim'$ to be the reverse ranking such that for each pair $A, B \in \mathcal{A}$, $A \succsim' B$ if and only if $B \succsim A$, (iii)~observe that $(\mathcal{A}, \supseteq, \succsim')$ is a smooth inference for which $(V_i)$ is descending, (iv)~apply the lemma to obtain the conclusion for $(\mathcal{A}, \supseteq, \succsim')$, and (v)~immediately obtain the conclusion for $(\mathcal{A}, \supseteq, \succsim)$.

Thus let us assume that $(V_i)$ is descending. In this case, we claim that each negative event contains a purely negative event. Indeed, let $A^- \in \mathcal{A}$ be negative.

Define $A_1 \equiv A^-$. For each $i \in \mathbb{N}$, we construct a mosaic $\mathcal{M}_i$ and an event $A_{i+1}$ as follows. First, define $A_{i, 1} \equiv A_i$. Second, for each $j \in \mathbb{N}$, (i)~if there is a subevent of $A_{i, j}$ in $\mathcal{UCS}(V_i)$, then select such an event $M_{i, j}$, (ii)~if there is no subevent of $A_{i, j}$ in $\mathcal{UCS}(V_i)$, then define $M_{i, j} \equiv \emptyset$, and (iii)~define $A_{i, j+1} \equiv A_{i, j} \backslash M_{i, j}$. Finally, define $\mathcal{M}_i \equiv \{M_{i, j}\}_{j \in \mathbb{N}} \backslash \{\emptyset\}$ and define $A_{i+1} \equiv A_i \backslash \cup \mathcal{M}_i$.

Define $\mathcal{M} \equiv \cup_{i \in \mathbb{N}} \mathcal{M}_i$. Since $\mathcal{M}$ is a purely positive mosaic, thus by \hyperlink{Lemma5}{Lemma~5} it is countable, so $\cup \mathcal{M} \in \mathcal{A}$, and moreover by \hyperlink{Lemma6}{Lemma~6} we have that $\cup \mathcal{M}$ is positive.

Define $A^\ominus \equiv A^- \backslash (\cup \mathcal{M})$. We cannot have $A^\ominus \succsim \emptyset$; else by {\it separability} we have $A^- = A^\ominus \cup (\cup\mathcal{M}) \succ \emptyset$, contradicting $\emptyset \succ A^-$. Moreover, assume by way of contradiction that $A^\ominus$ contains a positive event $A^+$. For each $i \in \mathbb{N}$, since $A^+ \subseteq A^\ominus = A^- \backslash (\cup \mathcal{M})$, thus $A^+ \subseteq A_{i+1}$ and $A^+ \not \in \mathcal{M}_i$, so by construction of $\mathcal{M}_i$ we have $A^+ \not \in \mathcal{UCS}(V_i)$, so $V_i \succ A^+$. But then by {\it continuity}, $\emptyset = \cap V_i \succsim A^+$, contradicting $A^+ \succ \emptyset$. Altogether, then, $A^\ominus$ is purely negative, as desired. Since $A^-$ was an arbitrary negative event, we are done.~$\blacksquare$

\hypertarget{AppendixD}{}
\section{Appendix D - Hahn decomposition for idealized spaces}

In this appendix, we prove \hyperlink{Proposition3}{Proposition~3}.

\vspace{\baselineskip} \noindent \textsc{Definition:} We say that an inference is {\it idealized} if and only if it has a unique purely null event.

\hypertarget{Proposition3}{}
\vspace{\baselineskip} \noindent \textsc{Proposition 3:} For each smooth and idealized inference, there is a Hahn decomposition.

\vspace{\baselineskip} \noindent \textsc{Proof:} Assume the hypotheses. First, we claim that $(\mathcal{A}, \supseteq)$ satisfies the {\it countable chain condition}: each antichain in $\mathcal{A}^*$ is countable.\footnote{Tragically, the established terminology for this countable antichain condition is indeed {\it the countable chain condition}.} Indeed, assume by way of contradiction that there is an uncountable antichain $\mathcal{A}^* \subseteq \mathcal{A}$. Let $\mathcal{A}^{**}$ be the collection constructed from $\mathcal{A}^*$ as follows: (i)~if $\emptyset \in \mathcal{A}^*$, then discard it; and (ii)~for each $[A] \in \mathcal{A}^*$ that is null but not purely null, select non-null $A' \subseteq A$ and then replace $A$ with $A'$. Since $\mathcal{A}^*$ is an uncountable antichain, and since we construct $\mathcal{A}^{**}$ from $\mathcal{A}^*$ by discarding at most one event and replacing some other events with associated subevents, thus $\mathcal{A}^{**}$ is an uncountable antichain. Moreover, (i)~since $(\mathcal{A}, \supseteq, \succsim)$ is idealized, thus $\emptyset$ is the unique purely null event, so since $\emptyset \not \in \mathcal{A}^{**}$ we have that $\mathcal{A}^{**}$ includes no purely null events, and (ii)~by construction, $\mathcal{A}^{**}$ includes no null event that is not purely null; thus $\mathcal{A}^{**}$ includes no null events. Altogether, then, $\mathcal{A}^{**}$ is an uncountable mosaic. But since $(\mathcal{A}, \supseteq, \succsim)$ is a smooth inference, thus by \hyperlink{Lemma5}{Lemma~5} we have that each of its mosaics is countable, contradicting that $\mathcal{A}^{**}$ is uncountable.

Second, we focus on the case that $\mathcal{A}$ has both a negative event and a positive event, as the result for this case implies the result. Indeed, if $\mathcal{A}$ has no negative event, then it follows from {\it separability} that we can take $H^\oplus = S$ and $H^\ominus = \emptyset$. Similarly, if $\mathcal{A}$ has no positive event, then it follows from {\it separability} that we can take $H^\oplus = \emptyset$ and $H^\ominus = S$.

Third, we focus on the case that each positive event contains a thoroughly positive event, as the result for this case implies the result. Indeed, by \hyperlink{Proposition2}{Proposition~2}, at least one of the following is true: (i)~each positive event contains a thoroughly positive event, or (ii)~each negative event contains a thoroughly negative event. Observe that if we establish the proposition supposing the former case holds, then we can easily prove the proposition supposing the latter case as follows: (i)~suppose the latter case holds, (ii)~define $\succsim'$ to be the reverse ranking such that for each pair $A, B \in \mathcal{A}$, $A \succsim' B$ if and only if $B \succsim A$, (iii)~observe that $(\mathcal{A}, \supseteq, \succsim')$ is a smooth and idealized inference with a thoroughly positive event, (iv)~apply the proposition for the former case to obtain the conclusion for $(\mathcal{A}, \supseteq, \succsim')$, and (v)~immediately obtain the conclusion for $(\mathcal{A}, \supseteq, \succsim)$.

Fourth, we construct $H^\oplus$ and $H^\ominus$. To begin, let $\mathcal{A}^\oplus \subseteq \mathcal{A}$ denote the collection of thoroughly positive events. Since there is a positive event, and since each positive event contains a thoroughly positive event, thus $\mathcal{A}^\oplus$ is nonempty. Moreover, since $(\mathcal{A}, \supseteq)$ satisfies the countable chain condition, thus by Lemma~1 of Section~14 in \cite{Halmos1963},\footnote{We remark that the proof of \cite{Halmos1963} uses Zorn's Lemma \citep{Zorn1935}, which is also used by \cite{Villegas1964} to partition each non-null event into two equivalent events.} there is countable $\mathcal{A}' \subseteq \mathcal{A}^\oplus$ such that $\mathcal{A}'$ and $\mathcal{A}^\oplus$ have the same upper bounds. Since $\mathcal{A}'$ is countable, this collection of upper bounds has a least member, and thus $\cup \mathcal{A}' = \cup \mathcal{A}^\oplus$. Since $\mathcal{A}'$ is a countable mosaic that is purely positive, thus by \hyperlink{Lemma6}{Lemma~6} we have $\cup \mathcal{A}^\oplus = \cup \mathcal{A}' \succ \emptyset$. Moreover, let $B \subseteq \cup \mathcal{A}^\oplus = \cup \mathcal{A}'$ and define $\mathcal{B} \equiv \{B \cap A | A \in \mathcal{A}'\}$; then $\mathcal{B}$ is a countable antichain consisting of subevents of thoroughly positive events, so $\mathcal{B}$ is a countable antichain of non-negative events, so by \hyperlink{Lemma6}{Lemma~6} we have $B = \cup \mathcal{B} \succsim \emptyset$. Since $B \subseteq \cup \mathcal{A}^\oplus$ was arbitrary, altogether $\cup \mathcal{A}^\oplus$ is a thoroughly positive event that contains all others. Define $H^\oplus = \cup \mathcal{A}^\oplus$ and define $H^\ominus \equiv S \backslash H^\oplus$.

Fifth, we claim that each subevent of $H^\oplus$ is non-negative and each subevent of $H^\ominus = S \backslash H^\oplus$ is non-positive. First, since $H^\oplus$ is throughly positive, thus each of its subevents is non-negative. Second, there cannot be a positive event $A \subseteq S \backslash H^\oplus$; else since each positive event contains a thoroughly positive event, there is a thoroughly positive event $A^\oplus$ that (i)~is contained in $\cup \mathcal{A}^\oplus = H^\oplus$ as a thoroughly positive event, and (ii)~is contained $A \subseteq S \backslash H^\oplus$, so $A^\oplus$ is contained in $H^\oplus \cap (S \backslash H^\oplus) = \emptyset$; but then $A^\oplus = \emptyset \sim \emptyset$, contradicting that $A^\oplus$ is positive.

To conclude, we verify that $H^\oplus$ and $H^\ominus$ satisfy the requirements. First, by the previous paragraph, for each $A \in \mathcal{A}$ we have that (i)~$H^\oplus \backslash A$ is non-negative, (ii)~$H^\oplus \cap A$ is non-negative, and (iii)~$A \backslash H^\oplus$ is non-positive; thus by {\it separability} we have that $H^\oplus = (H^\oplus \backslash A) \cup (H^\oplus \cap A) \succsim (H^\oplus \cap A) \succsim (H^\oplus \cap A) \cup (A \backslash H^\oplus) = A$, as desired. Second, by construction we have $H^\oplus \cap H^\ominus = \emptyset$ and $H^\oplus \cup H^\ominus = S$. Finally, by the previous paragraph we have the final two requirements.~$\blacksquare$

\hypertarget{AppendixE}{}
\section{Appendix E - Hahn decomposition for all spaces}

In this appendix, we prove \hyperlink{Proposition4}{Proposition~4}. To do, we first prove \hyperlink{Lemma9}{Lemma~9} and \hyperlink{Lemma10}{Lemma~10}; the former involves {\it $\sigma$-ideals} and the latter concerns {\it quotient spaces}.

\vspace{\baselineskip} \noindent \textsc{Definition:} Fix a $\sigma$-algebra $(\mathcal{A}, \supseteq)$. A collection of events $\mathcal{I} \subseteq \mathcal{A}$ is a {\it $\sigma$-ideal} if and only if (i) $\emptyset \in \mathcal{I}$, (ii) for each $A \in \mathcal{I}$ and each $B \in \mathcal{A}$ such that $B \subseteq A$, $B \in \mathcal{I}$, and (iii) for each countably infinite collection $\{A_i\}_{i \in \mathbb{N}} \subseteq \mathcal{I}$, we have $\cup A_i \in \mathcal{I}$.

\hypertarget{Lemma9}{}
\vspace{\baselineskip} \noindent \textsc{Lemma 9:} Fix a smooth inference. If $\mathcal{I} \subseteq \mathcal{A}$ is the collection of purely null events, then $\mathcal{I}$ is a {\it $\sigma$-ideal}.

\vspace{\baselineskip} \noindent \textsc{Proof:} Let $\mathcal{I}$ satisfy the hypothesis. By the definition purely null event, we have (i)~$\emptyset \in \mathcal{I}$, and (ii)~for each $A \in \mathcal{I}$ and each $B \in \mathcal{A}$ such that $B \subseteq A$, $B \in \mathcal{I}$. It remains to show that $\mathcal{I}$ is closed under countable suprema.

We first claim that if $A$ and $B$ are purely null, then $A \cup B$ is null. Indeed, in this case $A \backslash B$, $B \backslash A$, and $A \cap B$ are all null as subevents of purely null events, so by iterative application of {\it separability} we have $A \cup B \sim \emptyset$.

To conclude, let $\{A_i\}_{i \in \mathbb{N}} \subseteq \mathcal{I}$ and let $A' \subseteq \cup A_i$. For each $i \in \mathbb{N}$, define $B_i \equiv \cup_{j \leq i} (A' \cap A_j)$. For each $j \in \mathbb{N}$, $A_j$ is purely null, so $A' \cap A_j$ is purely null; thus by iterative application of the previous claim, $B_i$ is null. Since (i)~$B_1 \subseteq B_2 \subseteq ...$, and (ii)~for each $i \in \mathbb{N}$ we have $B_i \sim \emptyset$, thus by {\it continuity}, $A' = A' \cap (\cup A_i) = \cup (A' \cap A_i) = \cup B_i \sim \emptyset$. Since $A' \subseteq \cup A_i$ was arbitrary, thus $\cup A_i$ is purely null, so $\cup A_i \in \mathcal{I}$, as desired.~$\blacksquare$

\vspace{\baselineskip} We remark that under monotonicity, the collection of purely null events is simply the collection of null events, but without monotonicity, the collection of null events need not be a $\sigma$-ideal: it may be possible to partition a null event into a positive event and a negative event.

\vspace{\baselineskip} \noindent \textsc{Definition:} If $(\mathcal{A}, \supseteq, \succsim)$ is a smooth inference, and if $\mathcal{I} \subseteq \mathcal{A}$ is a $\sigma$-ideal, then the {\it $\mathcal{I}$-quotient of $(\mathcal{A}, \supseteq, \succsim)$} is the tuple $(\mathcal{A}^*, \supseteq^*, \succsim^*)$ defined as follows:
\begin{itemize}
\item For each $A \in \mathcal{A}$, $[A] \equiv \{B \in \mathcal{A} | (A \backslash B) \cup (B \backslash A) \in \mathcal{I}\}$; and $\mathcal{A}^* \equiv \{ [A] | A \in \mathcal{A} \}$. In other words, two events are declared equivalent if their symmetric difference belongs to $\mathcal{I}$; it is straightforward to verify that this is an equivalence relation because $\mathcal{I}$ is a $\sigma$-ideal; $\mathcal{A}^*$ is the collection of equivalence classes.

\item For each pair $[A], [B] \in \mathcal{A}^*$, $[A] \supseteq^* [B]$ if and only if there are $A' \in [A]$ and $B' \in [B]$ such that $A' \supseteq B'$.

\item For each pair $[A], [B] \in \mathcal{A}^*$, $[A] \succsim^* [B]$ if and only if there are $A' \in [A]$ and $B' \in [B]$ such that $A' \succsim B'$.
\end{itemize}
If moreover $\mathcal{I}$ is the collection of purely null events, then $(\mathcal{A}^*, \supseteq^*, \succsim^*)$ is the {\it null-quotient of $(\mathcal{A}, \supseteq, \succsim)$}. In this case, we emphasize that in order for two events to be declared equivalent, it is not enough for their symmetric difference to be null; their symmetric difference must moreover be purely null.

\hypertarget{Lemma10}{}
\vspace{\baselineskip} \noindent \textsc{Lemma 10:} Let $(\mathcal{A}, \supseteq, \succsim)$ be a smooth inference with null-quotient $(\mathcal{A}^*, \supseteq^*, \succsim^*)$. Let $\cup^*$, $\cap^*$, and $\neg^*$ denote supremum, infimum, and complement for $(\mathcal{A}^*, \supseteq^*)$, respectively. Then $(\mathcal{A}^*, \supseteq^*)$ is a $\sigma$-algebra such that
\begin{itemize}
\item for each pair $A, B \in \mathcal{A}$, $[A \cup B] = [A] \cup^* [B]$ and $[A \cap B] = [A] \cap^* [B]$

\item for each $A \in \mathcal{A}$, $[\neg A] = \neg^* [A]$,

\item for each $\{A_i\}_{i \in \mathbb{N}} \subseteq \mathcal{A}$, $[\cup A_i] = \cup^* [A_i]$, and

\item for each $A \in \mathcal{A}$, $[\emptyset] \subseteq^* [A] \subseteq^* [S]$.
\end{itemize}
In particular, $(\mathcal{A}^*, \supseteq^*)$ is a $\sigma$-algebra.

\vspace{\baselineskip} \noindent \textsc{Proof:} Assume the hypotheses. By \hyperlink{Lemma9}{Lemma~9}, the collection of purely null events is a $\sigma$-ideal, so by Section~13 of \cite{Halmos1963}, $(\mathcal{A}^*, \supseteq^*)$ is a $\sigma$-algebra that satisfies the first three conditions.\footnote{In the language of \cite{Halmos1963}, the projection $f: \mathcal{A} \to \mathcal{A}^*$ that maps each event $A$ to its equivalence class $[A]$ is a $\sigma$-homomorphism.} The final condition follows directly from the definition of null quotient: for each $A \in \mathcal{A}$, $\emptyset \subseteq A \subseteq S$, so $[\emptyset] \subseteq^* [A] \subseteq^* [S]$.~$\blacksquare$

\hypertarget{Proposition4}{}
\vspace{\baselineskip} \noindent \textsc{Proposition 4:} Let $(\mathcal{A}, \supseteq, \succsim)$ be a smooth inference with null-quotient $(\mathcal{A}^*, \supseteq^*, \succsim^*)$. Then $(\mathcal{A}^*, \supseteq^*, \succsim^*)$ is a smooth and idealized inference such that for each pair $A, B \in \mathcal{A}$, $A \succsim B$ if and only if $[A] \succsim^* [B]$.

\vspace{\baselineskip} \noindent \textsc{Proof:} Assume the hypotheses. We introduce the following notation: (i)~let $\mathcal{I} \subseteq \mathcal{A}$ denote the collection of purely null events in $\mathcal{A}$, (ii)~let $\cup^*$, $\cap^*$, and $\neg^*$ denote supremum, infimum, and complement for $(\mathcal{A}^*, \supseteq^*)$, respectively, (iii)~for each pair $[A], [B] \in \mathcal{A}^*$, let $[A] \backslash^* [B]$ denote $[A] \cap^* (\neg^* [B])$, and (iv)~define $\emptyset^* \equiv [\emptyset]$. For brevity, we will refer to properties (such as {\it order} and {\it separability}) without explicitly specifying the associated space---that is, $(\mathcal{A}, \supseteq, \succsim)$ or $(\mathcal{A}^*, \supseteq^*, \succsim^*)$---as this will always be clear from context. By \hyperlink{Lemma10}{Lemma~10}, $\emptyset^* = [\emptyset]$ is the minimum event in $(\mathcal{A}^*, \supseteq^*)$, and by definition it is the equivalence class of $\emptyset$; we use $\emptyset^*$ to emphasize the former and $[\emptyset]$ to emphasize the latter.

\vspace{\baselineskip} \noindent $\circ$ \textsc{Step 1:} For each pair $A, B \in \mathcal{A}$, (i)~$B \in [A]$ implies $A \sim B$, and (ii)~$A \succsim B$ if and only if $[A] \succsim^* [B]$. We use these facts freely throughout the rest of the proof.

\vspace{\baselineskip} For the first claim, let $A, B \in \mathcal{A}$ such that $B \in [A]$. Then $(A \backslash B) \cup (B \backslash A)$ is purely null, so $A \backslash B$ and $B \backslash A$ are null, so by {\it separability} we have $A = (A \cap B) \cup (A \backslash B) \sim A \cap B \sim (A \cap B) \cup (B \backslash A) = B$.

For the second claim, let $A, B \in \mathcal{A}$. If $A \succsim B$, then by construction $[A] \succsim^* [B]$. If $[A] \succsim^* [B]$, then there are $A' \in [A]$ and $B' \in [B]$ such that $A' \succsim B'$; thus by the previous claim we have $A \sim A' \succsim B' \sim B$.~$\square$

\vspace{\baselineskip} \noindent $\circ$ \textsc{Step 2:} $(\mathcal{A}^*, \supseteq^*, \succsim^*)$ is a smooth inference for which $\emptyset^*$ is the unique purely null event.

\vspace{\baselineskip} To see that $\emptyset^*$ is the unique purely null event in $\mathcal{A}^*$, let $[A] \in \mathcal{A}^*$ be purely null. For each $B \subseteq A$, $[B] \subseteq^* [A]$, so $[B]$ is a subevent of a purely null event, so $[B] \sim^* \emptyset^* = [\emptyset]$, so $B \sim \emptyset$. Since $B \subseteq A$ was arbitrary, thus $A = (A \backslash \emptyset) \cup (\emptyset \backslash A)$ is purely null, so altogether $[A] = [\emptyset] = \emptyset^*$, as desired.

To see that $(\mathcal{A}^*, \supseteq^*, \succsim^*)$ satisfies {\it order}, first let $[A], [B] \in \mathcal{A}^*$. Then by {\it order}, either $A \succsim B$ or $B \succsim A$, so either $[A] \succsim^* [B]$ or $[B] \succsim^* [A]$, as desired. Second, let $[A], [B], [C] \in \mathcal{A}^*$ such that $[A] \succsim^* [B]$ and $[B] \succsim^* [C]$. Then $A \succsim B$ and $B \succsim C$, so by {\it order}, $A \succsim C$, so $[A] \succsim^* [C]$.

To see that it satisfies {\it separability}, let $[A], [B], [C] \in \mathcal{A}^*$ such that $[A] \cap^* [C] = [B] \cap^* [C] = \emptyset^*$. Then by \hyperlink{Lemma10}{Lemma~10}, $[A \cap C] = [A] \cap^* [C] = \emptyset^* = [\emptyset]$ and $[B \cap C] = [B] \cap^* [C] = \emptyset^* = [\emptyset]$, so $A \cap C = ((A \cap C) \backslash \emptyset) \cup (\emptyset \backslash (A \cap C))$ is purely null and $B \cap C = ((B \cap C) \backslash \emptyset) \cup (\emptyset \backslash (B \cap C))$ is purely null; thus $(A \cap C) \backslash B$, $(B \cap C) \backslash A$, and $A \cap B \cap C$ are all null. Define $A' \equiv A \backslash C$, $B' \equiv B \backslash C$, and $C' \equiv C \backslash (A \cup B)$. It follows from repeated application of {\it separability} that $A \sim A'$, $B \sim B'$, $C \sim C \backslash A \sim C \backslash B \sim C'$, $A \cup C = A' \cup (A \cap C) \cup (C \backslash A) \sim A' \cup C'$, and $B \cup C = B' \cup (B \cap C) \cup (C \backslash B) \sim B' \cup C'$. Since $A' \cap C' = \emptyset$ and $B' \cap C' = \emptyset$, thus by {\it separability}, $A' \succsim B'$ if and only if $A' \cup C' \succsim B' \cup C'$; altogether, then, $A \succsim B$ if and only if $A \cup C \succsim B \cup C$. First, if $[A] \succsim^* [B]$, then $A \succsim B$, so $A \cup C \succsim B \cup C$, so $[A \cup C] \succsim^* [B \cup C]$, so by \hyperlink{Lemma10}{Lemma~10} we have $[A] \cup^* [C] \succsim^* [B] \cup^* [C]$. Second, if $[A] \cup^* [C] \succsim^* [B] \cup^* [C]$, then by \hyperlink{Lemma10}{Lemma~10} we have $[A \cup C] \succsim^* [B \cup C]$, so $A \cup C \succsim B \cup C$, so $A \succsim B$, so $[A] \succsim^* [B]$.

To see that it satisfies {\it continuity}, first let $[A] \in \mathcal{A}^*$ and let $([B_i]) \in (\mathcal{A}^*)^\mathbb{N}$ such that (i) $[B_1] \supseteq^* [B_2] \supseteq^* ...$, and (ii) for each $i \in \mathbb{N}$, $[B_i] \succsim^* [A]$. Then for each $i \in \mathbb{N}$, there are $B_i^i, B_{i+1}^i$ such that $B^i_i \in [B_i]$, $B^i_{i+1} \in [B_{i+1}]$, and $B^i_i \supseteq B^i_{i+1}$. For each $i \in \mathbb{N}$, define $N_i \equiv (B^{i+1}_{i+1} \backslash B^i_{i+1}) \cup (B^i_{i+1} \backslash B^{i+1}_{i+1})$; since $[B^i_{i+1}] = [B^{i+1}_{i+1}]$, thus $N_i$ is purely null. Define $N \equiv \cup N_i$; since $\{N_i\}_{i \in \mathbb{N}}$ is a countably infinite collection of purely null events, thus by \hyperlink{Lemma9}{Lemma~9} we have that $N$ is purely null. For each $i \in \mathbb{N}$, define $B'_i \equiv B^i_i \backslash N$; then (i)~since $N$ is purely null, thus $B^i_i \cap N = (B^i_i \backslash B'_i) \cup (B'_i \backslash B^i_i)$ is purely null, so $[B'_i] = [B^i_i] = [B_i]$, and (ii)~$[B_i] \succsim^* [A]$; thus $B'_i \sim B^i_i \sim B_i \succsim A$. Moreover, for each $i \in \mathbb{N}$,
\begin{align*}
B'_i &= B^i_i \backslash N
\\ &\supseteq B^i_{i+1} \backslash N
\\ &= (B^{i+1}_{i+1} \cap B^i_{i+1}) \backslash N
\\ &= B^{i+1}_{i+1} \backslash N
\\ &= B'_{i+1}.
\end{align*}
Thus $B'_1 \supseteq B'_2 \supseteq ...$, so by {\it continuity}, we have $\cap B'_i \succsim A$, so $[\cap B'_i] \succsim^* [A]$. Altogether, then, by \hyperlink{Lemma10}{Lemma~10}, we have $\cap^* [B_i] = \cap^* [B'_i] = [\cap B'_i] \succsim^* [A]$. Second, repeat the argument after replacing each instance of $\succsim^*$ and $\succsim$ with $\precsim^*$ and $\precsim$, respectively.

To see that it satisfies {\it no atoms}, let $[A] \in \mathcal{A}^*$ such that $[A] \not \sim^* \emptyset^*$. Then $A \not \sim \emptyset$, so by {\it no atoms}, there is $B \in \mathcal{A}$ such that $B \subseteq A$, $B \not \sim \emptyset$, and $B \not \sim A$. Thus $[B] \in \mathcal{A}^*$ satisfies $[B] \subseteq^* [A]$, $[B] \not \sim^* [\emptyset] = \emptyset^*$, and $[B] \not \sim^* [A]$, as desired.~$\blacksquare$

\hypertarget{Proposition5}{}
\vspace{\baselineskip} \noindent \textsc{Proposition 5:} For each smooth inference, there is a Hahn decomposition.

\vspace{\baselineskip} \noindent \textsc{Proof:} Assume the hypotheses. By \hyperlink{Proposition4}{Proposition~4}, the null-quotient is smooth and idealized, so by \hyperlink{Proposition3}{Proposition~3} it has a Hahn decomposition, so by \hyperlink{Lemma10}{Lemma~10} the original space has a Hahn decomposition.~$\blacksquare$ 

\hypertarget{AppendixF}{}
\section{Appendix F - Signed measure representation}

In this appendix, we prove \hyperlink{Theorem1}{Theorem~1}. To do so, we first prove \hyperlink{Lemma11}{Lemma~11} and \hyperlink{Lemma12}{Lemma~12}; the former concerns probability measure representations for {\it subspaces} and the latter concerns {\it annulments}.

\vspace{\baselineskip} \noindent \textsc{Definition:} Fix an assessment space. For each $A \in \mathcal{A}$, the {\it $A$-subspace} is the collection of subevents of $A$ together with the associated restrictions of the partial order and the assessment.

\hypertarget{Lemma11}{}
\vspace{\baselineskip} \noindent \textsc{Lemma 11:} Fix a smooth inference. For each Hahn decomposition $(H^\oplus, H^\ominus)$, if we let $(\mathcal{A}^\oplus, \supseteq^\oplus, \succsim^\oplus)$ denote $H^\oplus$-subspace and let $(\mathcal{A}^\ominus, \supseteq^\ominus, \succsim^\ominus)$ denote the $H^\ominus$-subspace, then
\begin{itemize}
\item $H^\oplus \succ \emptyset$ implies there is a unique $\mu^\oplus \in \mathbb{P}^\sigma_{\mathsf{na}}(\mathcal{A}^\oplus)$ such that $\mu^\oplus$ is a representation of $(\mathcal{A}^\oplus, \supseteq^\oplus, \succsim^\oplus)$, and

\item $\emptyset \succ H^\ominus$ implies there is a unique $\mu^\ominus \in \mathbb{P}^\sigma_{\mathsf{na}}(\mathcal{A}^\ominus)$ such that $-\mu^\ominus$ is a representation of $(\mathcal{A}^\ominus, \supseteq^\ominus, \succsim^\ominus)$.
\end{itemize}

\vspace{\baselineskip} \noindent \textsc{Proof:} Let $(H^\oplus, H^\ominus)$, $(\mathcal{A}^\oplus, \supseteq^\oplus, \succsim^\oplus)$, and $(\mathcal{A}^\ominus, \supseteq^\ominus, \succsim^\ominus)$ satisfy the hypotheses.

To begin, assume $H^\oplus \succ \emptyset$. First, since $(\mathcal{A}, \supseteq, \succsim)$ is a smooth inference, thus it is straightforward to verify that $(\mathcal{A}^\oplus, \supseteq^\oplus, \succsim^\oplus)$ is a smooth inference. Second, since each member of $\mathcal{A}^\oplus$ is a subevent of a thoroughly positive event, thus each member of $\mathcal{A}^\oplus$ is non-negative, so by {\it separability} we have that $(\mathcal{A}^\oplus, \supseteq^\oplus, \succsim^\oplus)$ satisfies {\it monotonicity}. Finally, since $H^\oplus \succ \emptyset$, thus $(\mathcal{A}^\oplus, \supseteq^\oplus, \succsim^\oplus)$ satisfies {\it non-degeneracy}. Altogether, then, we have that $(\mathcal{A}^\oplus, \supseteq^\oplus, \succsim^\oplus)$ is a qualitative probability that satisfies {\it continuity} and {\it no atoms}, so by \hyperlink{TheoremV2}{Theorem~V2} it has a unique representation $\mu^\oplus \in \mathbb{P}^\sigma_{\mathsf{na}}(\mathcal{A}^\oplus)$.

To conclude, assume $\emptyset \succ H^\ominus$. Let $\succsim'$ denote the reverse ranking such that for each pair $A, B \in \mathcal{A}$, $A \succsim' B$ if and only if $B \succsim A$, and let $\succsim''$ denote the restriction to $\mathcal{A}^\ominus$. Since $(\mathcal{A}, \supseteq, \succsim)$ is a smooth inference in which $H^\ominus$ is a thoroughly negative event, thus it is straightforward to verify that $(\mathcal{A}, \supseteq, \succsim')$ is a smooth inference in which $H^\ominus$ is a thoroughly positive event, so by the argument in the previous paragraph, $(\mathcal{A}, \supseteq, \succsim'')$ has a unique representation $\mu^\ominus \in \mathbb{P}^\sigma_{\mathsf{na}}(\mathcal{A}^\ominus)$. It follows directly that $\mu^\ominus$ is the unique member of $\mathbb{P}^\sigma_{\mathsf{na}}(\mathcal{A}^\ominus)$ such that $-\mu^\ominus$ is a representation of $(\mathcal{A}^\ominus, \supseteq^\ominus, \succsim^\ominus)$.~$\blacksquare$

\vspace{\baselineskip} \noindent \textsc{Definition:} Fix an assessment space. For each pair $A, B \in \mathcal{A}$, we say that $A$ is an {\it annulment of $B$} (and $B$ is an {\it annulment of $A$}) if and only if (i)~$A \cap B = \emptyset$, and (ii)~$A \cup B \sim \emptyset$.

\hypertarget{Lemma12}{}
\vspace{\baselineskip} \noindent \textsc{Lemma 12:} Fix a smooth inference. For each disjoint pair $A^+, A^- \in \mathcal{A}$ such that $A^+$ is purely non-negative and $A^-$ is purely non-positive.
\begin{itemize}
\item $A^+ \cup A^- \succsim \emptyset$ implies that $A^+$ contains an annulment of $A^-$, and

\item $A^+ \cup A^- \precsim \emptyset$ implies that $A^-$ contains an annulment of $A^+$.
\end{itemize}

\vspace{\baselineskip} \noindent \textsc{Proof:} Let $A^+$ and $A^-$ satisfy the hypotheses.

Observe that if we establish the lemma for the case that $A^+ \cup A^-$ is non-negative, then we can easily prove the lemma for the case that $A^+ \cup A^-$ is non-positive as follows: (i)~suppose that$A^+ \cup A^-$ is non-positive, (ii)~define $\succsim'$ to be the reverse ranking such that for each pair $A, B \in \mathcal{A}$, $A \succsim' B$ if and only if $B \succsim A$, (iii)~observe that $(\mathcal{A}, \supseteq, \succsim')$ is a smooth inference for which $A^+ \cup A^-$ is non-negative, (iv)~apply the lemma to obtain the conclusion for $(\mathcal{A}, \supseteq, \succsim')$, and (v)~immediately obtain the conclusion for $(\mathcal{A}, \supseteq, \succsim)$. Thus let us assume that $A^+ \cup A^- \succsim \emptyset$.

We focus on the case that $A^+ \succ \emptyset$, as the result for this case implies the result. Indeed, assume $A^+ \sim \emptyset$. In this case, we cannot have $A^- \prec \emptyset$; else by {\it separability} we have $A^+ \cup A^- \prec \emptyset$, contradicting $A^+ \cup A^- \succsim \emptyset$. Then since $A^-$ is purely non-positive, thus $A^- \sim \emptyset$, so by {\it separability} we have $A^+ \cup A^- \sim \emptyset$, so $A^+$ is an annulment of $A^-$ that is contained in $A^+$, as desired.

By \hyperlink{Proposition5}{Proposition~5}, there is a Hahn decomposition $(H^\oplus, H^\ominus)$. Let $(\mathcal{A}^\oplus, \supseteq^\oplus, \succsim^\oplus)$ denote the $H^\oplus$-subspace. Since $H^\oplus \succsim A^+ \succ \emptyset$, thus by \hyperlink{Lemma11}{Lemma~11}, $(\mathcal{A}^\oplus, \supseteq^\oplus, \succsim^\oplus)$ has a representation $\mu \in \mathbb{P}^\sigma_{\mathsf{na}}(\mathcal{A}^\oplus)$. Moreover, by \hyperlink{TheoremS}{Theorem~S} there is an $(A^+|\mu)$-knife, $\{A_v\}_{v \in [0, \mu(A^+)]}$. From here, the rest of the proof establishes that there is $v^* \in [0, \mu(A^+)]$ such that $A_{v^*}$ is an annulment of $A^-$.

Indeed, define $V^+ \equiv \{v \in [0, 1] | A_v \cup A^- \succsim \emptyset\}$ and $V^- \equiv \{v \in [0, 1] | A_v \cup A^- \precsim \emptyset\}$. We claim that $\mu(A^+) \in V^+$ and $0 \in V^-$. Indeed, since $A_{\mu(A^+)}$ and $A^+$ are both subevents of $A^+$, since $\mu$ represents comparisons of subevents of $A^+$, and since $\mu(A_{\mu(A^+)}) = \mu(A^+)$, thus $A_{\mu(A^+)} \sim A^+$, so by {\it separability} $A_{\mu(A^+)} \cup A^- \sim A^+ \cup A^- \sim \emptyset$ and thus $\mu(A^+) \in V^+$. Similarly, since $A_0$ and $\emptyset$ are both subevents of $A^+$, since $\mu$ represents comparisons of subevents of $A^+$, and since $\mu(A_0) = 0 = \mu(\emptyset)$, thus $A_0 \sim \emptyset$, so by {\it separability} $A_0 \cup A^- \sim \emptyset \cup A^- = A^- \prec \emptyset$ and thus $0 \in V^-$.

Define $v^+ \equiv \inf V^+$ and $v^- \equiv \sup V_-$; since $V^+$ and $V^-$ are nonempty subsets of $[0, \mu(A^+)]$, thus $v^+$ and $v^-$ belong to $[0, \mu(A^+)]$. We claim that $(v^+, \mu(A^+)] \subseteq V^+$ and $[0, v^-) \subseteq V^-$. Indeed, for each $v \in (v^+, \mu(A^+)]$, by definition of $v^+$ there is $v' \in V^+$ such that $v' \in (v^+, v)$, so $\mu(A_v) \geq \mu(A_{v'})$ and thus $A_v \succsim A_{v'}$, so by {\it separability} and the fact that $v' \in V^+$ we have $A_v \cup A^- \succsim A_{v'} \cup A^- \succsim \emptyset$, so $v \in V^+$. Similarly, for each $v \in [0, v^-)$, by definition of $v^-$ there is $v' \in V^-$ such that $v' \in (v, v^-)$, so $\mu(A_v) \leq \mu(A_{v'})$ and thus $A_v \precsim A_{v'}$, so by {\it separability} and the fact that $v' \in V^-$ we have $A_v \cup A^- \precsim A_{v'} \cup A^- \precsim \emptyset$, so $v \in V^-$.

We claim that $v^+ = v^-$. Indeed, we cannot have $v^- < v^+$; else there is $v \in (v^-, v^+)$, so $A_v \not \in V^+$ and $A_v \not \in V^-$, so neither $A_v \cup A^- \succsim \emptyset$ nor $A_v \cup A^- \precsim \emptyset$, contradicting that $\succsim$ is complete. Moreover, we cannot have $v^- > v^+$; else there are $v, v' \in (v^+, v^-)$ with $v' > v$, so (i)~$v, v' \in V^+ \cap V^-$ and thus $A_v \cup A^- \sim \emptyset \sim A_{v'} \cup A^-$, and (ii)~$\mu(A_{v'}) > \mu(A_v)$ and thus $A_{v'} \succ A_v$; but then by {\it separability} we have $\emptyset \sim A_{v'} \cup A^- \succ A_v \cup A^- \sim \emptyset$, contradicting $\emptyset \sim \emptyset$.

Define $v^* \equiv v^+ = v^-$. To conclude, we first claim that $A_{v^*} \cup A^- \succsim \emptyset$. Indeed, if $v^* = \mu(A^+)$, then $\mu(A_{v^*}) = v^* = \mu(A^+)$, so $A_{v^*} \sim A^+$, so by {\it separability} we have $A_{v^*} \cup A^- \sim A^+ \cup A^- \succsim \emptyset$, as desired. If $v^* < \mu(A^+)$, then there is a decreasing sequence $(v_i) \in (v^*, \mu(A^+)]^\mathbb{N}$ with limit $v^*$, and for each $i \in \mathbb{N}$ we have $v_i \in (v^*, \mu(A^+)] \subseteq V^+$ and thus $A_{v_i} \cup A^- \succsim \emptyset$, so by {\it continuity} we have $(\cap A_{v_i}) \cup A^- = \cap (A_{v_i} \cup A^-) \succsim \emptyset$, and moreover by the continuity of $\sigma$-additive measures we have $\mu(\cap A_{v_i}) = \lim v_i = v^* = \mu(A_{v^*})$ and thus $\cap A_{v_i} \sim A_{v^*}$, so altogether by {\it separability} we have $A_{v^*} \cup A^- \sim (\cap A_{v_i}) \cup A^- \succsim \emptyset$, as desired.

Similarly, we claim that $A_{v^*} \cup A^- \precsim \emptyset$. Indeed, if $v^* = 0$, then $\mu(A_{v^*}) = v^* = \mu(\emptyset)$, so $A_{v^*} \sim \emptyset$, so by {\it separability} we have $A_{v^*} \cup A^- \sim \emptyset \cup A^- = A^- \precsim \emptyset$, as desired. If $v^* > 0$, then there is an increasing sequence $(v_i) \in [0, v^*)^\mathbb{N}$ with limit $v^*$, and for each $i \in \mathbb{N}$ we have $v_i \in [0, v^*) \subseteq V^-$ and thus $A_{v_i} \cup A^- \precsim \emptyset$, so by {\it continuity} we have $(\cup A_{v_i}) \cup A^- = \cup (A_{v_i} \cup A^-) \precsim \emptyset$, and moreover by the continuity of $\sigma$-additive measures we have $\mu(\cup A_{v_i}) = \lim v_i = v^* = \mu(A_{v^*})$ and thus $\cup A_{v_i} \sim A_{v^*}$, so altogether by {\it separability} we have $A_{v^*} \cup A^- \sim (\cup A_{v_i}) \cup A^- \precsim \emptyset$, as desired.

By the previous two paragraphs, $A_{v^*}$ is a subevent of $A^+$ with the property that $A_{v^*} \cup A^- \sim \emptyset$; thus $A^+$ contains $A_{v^*}$, which is an annulment of $A^-$, as desired.~$\blacksquare$

\vspace{\baselineskip} \noindent \textsc{Theorem 1:} A {\it non-degenerate} inference is smooth if and only if it has a normalized representation $\mu \in \mathbb{S}^\sigma_{\mathsf{na}}(\mathcal{A})$. In this case, there is no other normalized representation in all of $\mathbb{S}(\mathcal{A}) \supseteq \mathbb{S}^\sigma_{\mathsf{na}}(\mathcal{A})$; $\mu$ is the unique normalized signed measure representation.

\vspace{\baselineskip} \noindent \textsc{Proof:} Fix an assessment space. It is straightforward to show that if a {\it non-degenerate} inference has a normalized signed measure representation, then it is smooth; we omit the argument. Thus let us assume we have a {\it non-degenerate} inference that is smooth.

By \hyperlink{Proposition5}{Proposition~5}, there is a Hahn decomposition $(H^\oplus, H^\ominus)$. Let $(\mathcal{A}^\oplus, \supseteq^\oplus, \succsim^\oplus)$ denote the $H^\oplus$-subspace and let $(\mathcal{A}^\ominus, \supseteq^\ominus, \succsim^\ominus)$ denote the $H^\ominus$-subspace.

\vspace{\baselineskip} \noindent $\circ$ \textsc{Step 1:} Restrict attention the case that $H^\oplus \succ \emptyset \succ H^\ominus$ and $S = H^\oplus \cup H^\ominus \succsim \emptyset$.

\vspace{\baselineskip} To begin, we restrict attention to the case that $H^\oplus \succ \emptyset \succ H^\ominus$ by handling the other cases. Indeed, we cannot have $H^\oplus \sim \emptyset$ and $H^\ominus \sim \emptyset$, as in this case {\it non-degeneracy} is violated. If $H^\oplus \succ \emptyset$ and $H^\ominus \sim \emptyset$, then since each event is non-negative, thus by {\it separability} we have {\it monotonicity}, so $(\mathcal{A}, \supseteq, \succsim)$ is a qualitative probability that satisfies {\it continuity} and {\it no atoms}, so by \hyperlink{TheoremV2}{Theorem~V2} we have that $(\mathcal{A}, \supseteq, \succsim)$ has a unique representation $\mu$ in $\mathbb{P}^\sigma_{\mathsf{na}}(\mathcal{A})$, from which the desired conclusion follows from \hyperlink{Proposition1}{Proposition~1} and {\it no atoms}. If $H^\oplus \sim \emptyset$ and $\emptyset \succ H^\ominus$, then let $\succsim'$ denote the reverse ranking such that for each pair $A, B \in \mathcal{A}$, $A \succsim' B$ if and only if $B \succsim A$; by the previous argument, $(\mathcal{A}, \supseteq, \succsim')$ has a unique representation $\mu$ in $\mathbb{P}^\sigma_{\mathsf{na}}(\mathcal{A})$, so $-\mu$ is the unique representation of $(\mathcal{A}, \supseteq, \succsim)$ in $\mathbb{S}^\sigma_{\mathsf{na}}(\mathcal{A})$, from which the desired conclusion follows from \hyperlink{Proposition1}{Proposition~1} and {\it no atoms}.

Observe that if we establish the theorem for the case that $S$ is non-negative, then we can easily prove the lemma for the case that $S$ is non-positive as follows: (i)~suppose that $S$ is non-positive, (ii)~define $\succsim'$ to be the reverse ranking such that for each pair $A, B \in \mathcal{A}$, $A \succsim' B$ if and only if $B \succsim A$, (iii)~observe that $(\mathcal{A}, \supseteq, \succsim')$ is a smooth inference for which $S$ is non-negative, (iv)~apply the theorem to obtain the conclusion for $(\mathcal{A}, \supseteq, \succsim')$, and (v)~immediately obtain the conclusion for $(\mathcal{A}, \supseteq, \succsim)$. Thus let us assume that $S = H^\oplus \cup H^\ominus$ is non-negative.~$\square$

\vspace{\baselineskip} \noindent $\circ$ \textsc{Step 2:} Define $\mu^\oplus$, $\mu^\ominus$, $s^\ominus$, and $\mu$.

\vspace{\baselineskip} Since $H^\oplus \succ \emptyset \succ H^\ominus$, thus by \hyperlink{Lemma11}{Lemma~11}, (i)~there is a unique $\mu^\oplus \in \mathbb{P}^\sigma_{\mathsf{na}}(\mathcal{A}^\oplus)$ such that $\mu^\oplus$ is a representation of $(\mathcal{A}^\oplus, \supseteq^\oplus, \succsim^\oplus)$, and (ii)~there is a unique $\mu^\ominus \in \mathbb{P}^\sigma_{\mathsf{na}}(\mathcal{A}^\ominus)$ such that $-\mu^\ominus$ is a representation of $(\mathcal{A}^\ominus, \supseteq^\ominus, \succsim^\ominus)$.

Let $\mathcal{A}^* \subseteq \mathcal{A}$ denote the collection of annulments of $H^\ominus$ contained in $H^\oplus$. Since (i)~$H^\oplus$ and $H^\ominus$ are disjoint, (ii)~$H^\oplus$ is purely non-negative, (iii)~$H^\ominus$ is purely non-positive, and (iv)~$H^\oplus \cup H^\ominus \succsim \emptyset$, thus by \hyperlink{Lemma12}{Lemma~12} we have that $\mathcal{A}^*$ is nonempty. For each pair $A, B \in \mathcal{A}^*$, we have $H^\ominus \cap A = \emptyset = H^\ominus \cap B = \emptyset$ and $H^\ominus \cup A \sim \emptyset \sim H^\ominus \cup B$, so by {\it separability} we have $A \sim B$; thus we can define $s^\ominus \in [0, 1]$ to be the unique member of $\{\mu^\oplus(A) | A \in \mathcal{A}^*\}$. The notation $s^\ominus$ is intended to suggest a new scale for $\mu^\ominus$, and indeed, we define $\mu: \mathcal{A} \to [0, 1]$ as follows: for each $A \in \mathcal{A}$, $\mu(A) \equiv \mu^\oplus(A \cap H^\oplus) - s^\ominus \cdot \mu^\ominus(A \cap H^\ominus)$.~$\square$

\vspace{\baselineskip} \noindent $\circ$ \textsc{Step 3:} Prove that $\mu \in \mathbb{S}^\sigma_{\mathsf{na}}(\mathcal{A})$ with $\sup_{A \in \mathcal{A}} |\mu(A)| = \max_{A \in \mathcal{A}} |\mu(A)| = 1$.

\vspace{\baselineskip} First, we claim that $\mu$ is countably additive. Indeed, let $(A_i) \in \mathcal{A}^\mathbb{N}$ such that $\{A_i\}_{i \in \mathbb{N}}$ is pairwise-disjoint. For each $i \in \mathbb{N}$, define $A^+_i \equiv A_i \cap H^\oplus$ and $A^-_i \equiv A_i \cap H^\ominus$. Since $\mu^\oplus$ and $\mu^\ominus$ are countably additive, thus
\begin{align*}
\mu(\cup A_i) &= \mu^\oplus((\cup A_i) \cap H^\oplus) - s^\ominus \cdot \mu^\ominus((\cup A_i) \cap H^\ominus)
\\ &= \mu^\oplus(\cup A^+_i) - s^\ominus \cdot \mu^\ominus(\cup A^-_i)
\\ &= \sum \mu^\oplus(A^+_i) - s^\ominus \cdot \sum \mu^\ominus(A^-_i)
\\ &= \sum (\mu^\oplus(A^+_i) - s^\ominus \cdot \mu^\ominus(A^-_i))
\\ &= \sum \mu(A_i).
\end{align*}
Since $(A_i) \in \mathcal{A}^\mathbb{N}$ with $\{A_i\}$ pairwise-disjoint was arbitrary, thus $\mu$ is countably additive, as desired.

Second, we claim that $\mu$ has no measure-atoms. Indeed, let $A \in \mathcal{A}$ such that $\mu(A) \neq 0$, define $A^+ \equiv A \cap H^\oplus$, and define $A^- \equiv A \cap H^\ominus$. If $\mu^\oplus(A^+) > 0$, then since $\mu^\oplus \in \mathbb{P}^\sigma_{\mathsf{na}}(\mathcal{A}^\oplus)$, thus by \hyperlink{TheoremS}{Theorem~S} there is $B \subseteq A^+$ such that $\mu^\oplus(B) \not \in \{\mu^\oplus(A^+), s^\ominus \cdot \mu^\ominus(A^-)\}$, and it is straightforward to show that $B \cup A^-$ is a subevent of $A$ such that $\mu(B \cup A^-) \not \in \{0, \mu(A)\}$. If $\mu^\oplus(A^+) \leq 0$, then since $\mu(A) \neq 0$ necessarily $\mu^\ominus(A^-) < 0$, so since $\mu^\ominus \in \mathbb{P}^\sigma_{\mathsf{na}}(\mathcal{A}^\ominus)$, thus by \hyperlink{TheoremS}{Theorem~S} there is $B \subseteq A^-$ such that $\mu^\ominus(B) \not \in \{\mu^\ominus(A^-), \frac{1}{s^\ominus} \cdot \mu^\oplus(A^+)\}$, and it is straightforward to show that $A^+ \cup B$ is a subevent of $A$ such that $\mu(A^+ \cup B) \not \in \{0, \mu(A)\}$.

Finally, we claim that $\sup_{A \in \mathcal{A}} |\mu(A)| = \max_{A \in \mathcal{A}} |\mu(A)| = 1$. Indeed, $\mu(H^\oplus) = 1$. Moreover, for each $A \in \mathcal{A}$, we have $\mu^\oplus(A \cap H^\oplus) \in [0, 1]$ and $s^\ominus \cdot \mu^\ominus(A \cap H^\ominus) \in [0, s^\ominus] \subseteq [0, 1]$, so $\mu(A) \in [-1, 1]$. Altogether, then, $\sup_{A \in \mathcal{A}} |\mu(A)| = \max_{A \in \mathcal{A}} |\mu(A)| = 1$, as desired.~$\square$

\vspace{\baselineskip} \noindent $\circ$ \textsc{Step 4:} For each $A^+ \in \mathcal{A}^\oplus$ and each $A^- \in \mathcal{A}^\ominus$, $\mu(A^+) = -\mu(A^-)$ implies $A^+ \cup A^- \sim \emptyset$.

\vspace{\baselineskip} Define $V \equiv \{v \in [0, s^\ominus]|A^+ \in \mathcal{A}^\oplus, A^- \in \mathcal{A}^\ominus, \mu(A^+) = -\mu(A^-) = v \Rightarrow A^+ \cup A^- \sim \emptyset\}$. Since for each $A^- \in \mathcal{A}^\ominus$, we have $-\mu(A^-) \in [0, s^\ominus]$, thus it suffices to show $V = [0, s^\ominus]$.

First, we claim $s^\ominus \in V$. Indeed, let $A^+ \in \mathcal{A}^\oplus$ and $A^- \in \mathcal{A}^\ominus$ such that $\mu(A^+) = -\mu(A^-) = s^\ominus$. Since we have $s^\ominus = -\mu(A^-) = s^\ominus \cdot \mu^\ominus(A^-)$, thus $\mu^\ominus(H^\ominus) = 1 = \mu^\ominus(A^-)$, so $H^\ominus \sim A^-$. Moreover, by the construction of $\mu$ in Step~2, there is $B^+ \in \mathcal{A}$ such that (i)~$B^+$ is an annulment of $H^\ominus$ contained in $H^\oplus$, and (ii)~$\mu^\oplus(B^+) = s^\ominus$. Since $\mu^\oplus(B^+) = s^\ominus = \mu^\oplus(A^+)$, thus $B^+ \sim A^+$. Altogether, then, since $B^+ \cup H^\ominus \sim \emptyset$, thus by {\it separability} we have $B^+ \cup A^- \sim \emptyset$, so by {\it separability} again we have $A^+ \cup A^- \sim \emptyset$, as desired. Since $A^+$ and $A^-$ satisfying the hypotheses were arbitrary, thus $s^\ominus \in V$.

Second, we claim that for each $b \in \mathbb{N}$ and each $a \in \{0, 1, ..., b\}$, $\frac{a}{b} \cdot s^\ominus \in V$. Indeed, let $b \in \mathbb{N}$, let $a \in \{0, 1, ..., b\}$, and define $v \equiv \frac{a}{b} \cdot s^\ominus$. Moreover, let $A^+ \in \mathcal{A}^\oplus$ and $A^- \in \mathcal{A}^\ominus$ such that $\mu(A^+) = -\mu(A^-) = v$. First, by \hyperlink{TheoremS}{Theorem~S}, there is $B^+ \subseteq H^\oplus$ such that $B^+$ contains $A^+$ and $\mu^\oplus(B^+) = s^\ominus$. Second, by \hyperlink{TheoremS}{Theorem~S}, there is a partition of $B^+$, $\{B^+_1, B^+_2, ..., B^+_b\}$, such that (i)~for each $i \in \{1, 2, ..., b\}$, $\mu^\oplus(B^+_i) = \frac{1}{b} \cdot \mu^\oplus(B^+)$, and (ii)~$A^+ = \cup_{i =1}^a B^+_i$. Third, by \hyperlink{TheoremS}{Theorem~S}, there is a partition of $H^\ominus$, $\{B^-_1, B^-_2, ..., B^-_b\}$, such that (i)~for each $i \in \{1, 2, ..., b\}$, $\mu^\ominus(B^-_i) = \frac{1}{b} \cdot \mu^\oplus(H^\ominus)$, and (ii)~$A^- = \cup_{i =1}^a B^-_i$. Since $\mu^\oplus(B^+) = s^\ominus = - \mu^\ominus(H^\ominus)$, thus by the previous paragraph we have $\emptyset \sim B^+ \cup H^\ominus = \cup_{i \in \{1, 2, ..., b\}} (B^+_i \cup B^-_i)$. Moreover, for each pair $i, j \in \{1, 2, ..., b\}$, we have $B^+_i \sim B^+_j$ and $B^-_i \sim B^-_j$, so by {\it separability} we have $B^+_i \cup B^-_i \sim B^+_j \cup B^+_j$. Then it must be that for each $i \in \{1, 2, ..., b\}$ we have $B^+_i \cup B^-_i \sim \emptyset$; else either all are positive and all are negative, so by {\it separability} either $B^+ \cup H^\ominus$ is positive or $B^+ \cup H^\ominus$ is negative, contradicting that $B^+ \cup H^\ominus$ is null. Thus by {\it separability}, we have $A^+ \cup A^- = (\cup_{i = 1}^a B^+_i) \cup (\cup_{i =1}^a B^-_i) = \cup_{i = 1}^a (B^+_i \cup B^-_i) \sim \emptyset$, as desired. Since $A^+$ and $A^-$ satisfying the hypotheses were arbitrary, thus $v \in V$. Since $v$ satisfying the hypothesis was arbitrary, we have established the claim.

To conclude, let $v \in [0, s^\ominus]$. We claim that $v \in V$. If $v \in \{0, s^\ominus\}$, then by the previous paragraphs we are done; thus let us assume $v \not \in \{0, s^\ominus\}$. Let $A^+ \in \mathcal{A}^\oplus$ and $A^- \in \mathcal{A}^\ominus$ such that $\mu(A^+) = -\mu(A^-) = v$. Let $\mu'$ denote the restriction of $-\mu$ to $\mathcal{A}^\ominus$; this is simply $s^\ominus \cdot \mu^\ominus$. By \hyperlink{TheoremS}{Theorem~S}, there is an $(H^\ominus|\mu^\ominus)$-knife; thus there is an $(H^\ominus|\mu')$-knife, $(\kappa_v)_{v \in [0, s^\ominus]}$. Select $(v^\downarrow_i) \in [0, s^\ominus]^\mathbb{N}$ such that (i)~$v^\downarrow_1 > v^\downarrow_2 > ...$, (ii)~for each $i \in \mathbb{N}$, there is a rational number $q_i \in \mathbb{Q}$ such that $v^\downarrow_i = q_i \cdot s^\ominus$, and (iii)~$\lim v^\downarrow_i = v$. Similarly, select $(v^\uparrow_i) \in [0, s^\ominus]^\mathbb{N}$ such that (i)~$v^\uparrow_1 < v^\uparrow_2 < ...$, (ii)~for each $i \in \mathbb{N}$, there is a rational number $q_i \in \mathbb{Q}$ such that $v^\uparrow_i = q_i \cdot s^\ominus$, and (iii)~$\lim v^\uparrow_i = v$. We can indeed select such sequences because the rationals are dense in $[0, 1]$ and $v \in (0, s^\ominus)$. For each $i \in \mathbb{N}$, by \hyperlink{TheoremS}{Theorem~S} there are $B^\downarrow_i, B^\uparrow_i \in \mathcal{A}^\oplus$ such that $\mu(B^\downarrow_i) = \mu^\oplus(B^\downarrow_i) = v^\downarrow_i = \mu'(\kappa_{v^\downarrow_i}) = -\mu(\kappa_{v^\downarrow_i})$ and $\mu(B^\uparrow_i) = \mu^\oplus(B^\uparrow_i) = v^\uparrow_i = \mu'(\kappa_{v^\uparrow_i}) = -\mu(\kappa_{v^\uparrow_i})$. Then for each $i \in \mathbb{N}$, (i)~since $v^\downarrow_i > v > v^\uparrow_i$ and since $\mu^\oplus$ represents $(\mathcal{A}^\oplus, \supseteq^\oplus, \succsim^\oplus)$, we have $B^\downarrow_i \succ A^+ \succ B^\uparrow_i$, and (ii)~by the previous paragraph, we have $B^\downarrow_i \cup \kappa_{v^\downarrow_i} \sim \emptyset$ and $B^\uparrow_i \cup \kappa_{v^\uparrow_i} \sim \emptyset$; thus by {\it separability} we have $\emptyset \sim B^\downarrow_i \cup \kappa_{v^\downarrow_i} \succ A^+ \cup \kappa_{v^\downarrow_i}$ and $A^+ \cup \kappa_{v^\uparrow_i} \succ B^\uparrow_i \cup \kappa_{v^\uparrow_i} \sim \emptyset$. Then by {\it continuity}, $\emptyset \succsim A^+ \cup (\cap \kappa^\downarrow_v)$ and $A^+ \cup (\cup \kappa^\uparrow_v) \succsim \emptyset$. Moreover, since $\mu^\ominus$ is non-negative and countably additive, thus $\mu^\ominus$ is continuous in the topology of sequential order convergence,\footnote{This is well-known; see 326K in Chapter~32 of \cite{Fremlin2012}. For a complete proof of this well-known fact, see Lemma~2 in \cite{Mackenzie2019}. For more on this topology, see \cite{Maharam1947}, \cite{Vladimirov2002}, and \cite{Balcar-Jech-Pazak2005}.} so $\mu'$ is continuous in the topology of sequential order convergence, so $\mu'(\cap \kappa^\downarrow_i) = \lim \mu'(\kappa^\downarrow_v) = \lim v^\downarrow_i = v = \lim v^\uparrow_i = \lim \mu'(\kappa^\uparrow_v) = \mu'(\cup \kappa^\uparrow_i)$. Thus $\mu'(A^-) = v = \mu'(\cap \kappa^\downarrow_i) = \mu'(\cup \kappa^\uparrow_i)$, so $\mu^\ominus(A^-) = \mu^\ominus(\cap \kappa^\downarrow_i) = \mu^\ominus(\cup \kappa^\uparrow_i)$, so since $-\mu^\ominus$ represents $(\mathcal{A}^\ominus, \supseteq^\ominus, \succsim^\ominus)$, we have $A^- \sim \cap \kappa^\downarrow_i \sim \cup \kappa^\uparrow_i$. Altogether, then, by {\it separability} we have $\emptyset \succsim A^+ \cup A^-$ and $A^+ \cup A^- \succsim \emptyset$, so $A^+ \cup A^- \sim \emptyset$, as desired. Since $A^+$ and $A^-$ satisfying the hypotheses were arbitrary, thus $v \in V$. Since $v$ satisfying the hypothesis was arbitrary, we are done.~$\square$

\vspace{\baselineskip} \noindent $\circ$ \textsc{Step 5:} For each $A^+ \in \mathcal{A}^\oplus$ and each $A^- \in \mathcal{A}^\ominus$, $A^+ \cup A^- \sim \emptyset$ implies $\mu(A^+) = -\mu(A^-)$.

\vspace{\baselineskip} Let $A^+$ and $A^-$ satisfy the hypotheses.

First, assume by way of contradiction that $\mu(A^+) > -\mu(A^-)$. Then $\mu^\oplus(A^+) > -\mu(A^-)$, so by \hyperlink{TheoremS}{Theorem~S}, there is $B \in \mathcal{A}^\oplus$ such that $\mu^\oplus(A^+) > -\mu(A^-) = \mu^\oplus(B)$. Since $\mu^\oplus$ is a representation of $(\mathcal{A}^\oplus, \supseteq^\oplus, \succsim^\oplus)$, thus $A^+ \succ B$. Moreover, $\mu(B) = \mu^\oplus(B) = -\mu(A^-)$, so by Step~4 we have $B \cup A^- \sim \emptyset$. But then by {\it separability} we have $A^+ \cup A^- \succ \emptyset$, contradicting $A^+ \cup A^- \sim \emptyset$.

Second, assume by way of contradiction that $\mu(A^+) < -\mu(A^-)$. Then $\frac{1}{s^\ominus} \cdot \mu(A^+) < \mu^\ominus(A^-)$, so by \hyperlink{TheoremS}{Theorem~S}, there is $B \in \mathcal{A}^\ominus$ such that $\mu^\ominus(B) = \frac{1}{s^\ominus} \cdot \mu(A^+) < \mu^\ominus(A^-)$. Since $-\mu^\oplus$ is a representation of $(\mathcal{A}^\ominus, \supseteq^\ominus, \succsim^\ominus)$, thus $A^- \prec B$. Moreover, $\mu(A^+) = s^\ominus \cdot \mu^\ominus(B) = -\mu(B)$, so by Step~4 we have $A^+ \cup B \sim \emptyset$. But then by {\it separability} we have $A^+ \cup A^- \prec \emptyset$, contradicting $A^+ \cup A^- \sim \emptyset$.~$\square$

\vspace{\baselineskip} \noindent $\circ$ \textsc{Step 6:} Prove that $\mu$ is a representation.

\vspace{\baselineskip} First, we claim that for each $C \in \mathcal{A}$ such that $C \succsim \emptyset$, there is $C' \in \mathcal{A}$ such that $C' \subseteq H^\oplus$, $C \sim C'$, and $\mu(C) = \mu(C')$. Indeed, define $C^+ \equiv C \cap H^\oplus$ and $C^- \equiv C \cap H^\ominus$. Since (i)~$C^+$ and $C^-$ are disjoint, (ii)~$C^+$ is purely non-negative, (iii)~$C^-$ is purely non-positive, and (iv)~$C^+ \cup C^- = C \succsim \emptyset$, thus by \hyperlink{Lemma12}{Lemma~12}, $C^+$ contains an annulment of $C^-$. Select such an annulment $\alpha(C^-)$, and define $C' \equiv C^+ \backslash \alpha(C^-)$; clearly $C' \subseteq H^\oplus$. Moreover, since $C^- \cup \alpha(C^-) \sim \emptyset$, thus by {\it separability} we have $C = C^+ \cup C^- = (C^+ \backslash \alpha(C^-)) \cup (C^- \cup \alpha(C^-)) \sim C^+ \backslash \alpha(C^-) = C'$, as desired. Finally, by Step~5 we have $\mu(\alpha(C^-)) = -\mu(C^-)$, so $\mu(C) = \mu(C^+ \cup C^-) = \mu((C^+ \backslash \alpha(C^-)) \cup (C^- \cup \alpha(C^-))) = \mu(C^+ \backslash \alpha(C^-)) + \mu(C^- \cup \alpha(C^-)) = \mu(C') + \mu(C^-) + \mu(\alpha(C^-)) = \mu(C')$, as desired.

Second, we claim that for each $C \in \mathcal{A}$ such that $C \precsim \emptyset$, there is $C' \in \mathcal{A}$ such that $C' \subseteq H^\ominus$, $C \sim C'$, and $\mu(C) = \mu(C')$. Indeed, define $C^+ \equiv C \cap H^\oplus$ and $C^- \equiv C \cap H^\ominus$. Since (i)~$C^+$ and $C^-$ are disjoint, (ii)~$C^+$ is purely non-negative, (iii)~$C^-$ is purely non-positive, and (iv)~$C^+ \cup C^- = C \precsim \emptyset$, thus by \hyperlink{Lemma12}{Lemma~12}, $C^-$ contains an annulment of $C^+$. Select such an annulment $\alpha(C^+)$, and define $C' \equiv C^- \backslash \alpha(C^+)$; clearly $C' \subseteq H^\ominus$. Moreover, since $C^+ \cup \alpha(C^+) \sim \emptyset$, thus by {\it separability} we have $C = C^+ \cup C^- = (C^- \backslash \alpha(C^+)) \cup (C^+ \cup \alpha(C^+)) \sim C^- \backslash \alpha(C^+) = C'$, as desired. Finally, by Step~5 we have $\mu(\alpha(C^+)) = -\mu(C^+)$, so $\mu(C) = \mu(C^+ \cup C^-) = \mu((C^- \backslash \alpha(C^+)) \cup (C^+ \cup \alpha(C^+))) = \mu(C^- \backslash \alpha(C^+)) + \mu(C^+ \cup \alpha(C^+)) = \mu(C') + \mu(C^+) + \mu(\alpha(C^+)) = \mu(C')$, as desired.

To conclude, let $A, B \in \mathcal{A}$. If $A \succsim \emptyset$, then by the first claim, there is $A' \in \mathcal{A}^\oplus$ such that $A \sim A'$ and $\mu(A) = \mu(A')$; else by the second claim, there is $A' \in \mathcal{A}^\ominus$ such that $A \sim A'$ and $\mu(A) = \mu(A')$. Similarly, if $B \succsim \emptyset$, then by the first claim, there is $B' \in \mathcal{A}^\oplus$ such that $B \sim B'$ and $\mu(B) = \mu(B')$; else by the second claim, there is $B' \in \mathcal{A}^\ominus$ such that $B \sim B'$ and $\mu(B) = \mu(B')$. We obtain the desired conclusion in each of four cases. First, if $A \succsim \emptyset$ and $B \succsim \emptyset$, then $A \succsim B$ if and only if $A' \succsim B'$ if and only if $\mu^\oplus(A') \geq \mu^\oplus(B')$ if and only if $\mu(A) \geq \mu(B)$. Second, if $A \succsim \emptyset$ and $\emptyset \succsim B$, then $A \succsim \emptyset \succsim B$ and $\mu(A) \geq 0 \geq \mu(B)$. Third, if $\emptyset \succsim A$ and $B \succsim \emptyset$, then $B \succsim \emptyset \succsim A$ and $\mu(B) \geq 0 \geq \mu(A)$. Finally, if $\emptyset \succsim A$ and $\emptyset \succsim B$, then $A \succsim B$ if and only if $A' \succsim B'$ if and only if $-\mu^\ominus(A') \geq -\mu^\ominus(B')$ if and only if $\mu(A') \geq \mu(B')$.~$\square$

\vspace{\baselineskip} \noindent $\circ$ \textsc{Step 7:} Conclude.

\vspace{\baselineskip} By Step~3, $\mu \in \mathbb{S}^\sigma_{\mathsf{na}}(\mathcal{A})$ with $\sup_{A \in \mathcal{A}} |\mu(A)| = \max_{A \in \mathcal{A}} |\mu(A)| = 1$, and by Step~6, $\mu$ is a normalized representation. To conclude, let $\mu' \in \mathbb{S}(\mathcal{A})$ be another normalized representation. By \hyperlink{Proposition1}{Proposition~1} and {\it no atoms}, $\mu \in \mathbb{S}^\sigma_{\mathsf{na}}(\mathcal{A})$.

First, we claim $\mu'(H^\oplus) = 1$. Indeed, since $H^\oplus \cup H^\ominus \succsim \emptyset$, thus $\mu'(H^\oplus) + \mu'(H^\ominus) = \mu'(H^\oplus \cup H^\ominus) \geq \mu'(\emptyset) = 0$, so $\mu'(H^\ominus) \geq -\mu'(H^\ominus)$. Moreover, for each $A \in \mathcal{A}$ we have $H^\oplus \succsim A \succsim H^\ominus$, so $\mu'(H^\oplus) \geq \mu'(A) \geq \mu'(H^\ominus)$, and in particular since $\mu'(\emptyset) = 0$ we have $\mu'(H^\oplus) \geq 0 \geq \mu'(H^\ominus)$. Thus since $\sup_{A \in \mathcal{A}} |\mu'(A)| = \max_{A \in \mathcal{A}} |\mu'(A)| = 1$, we have $\mu'(H^\oplus) = 1$, as desired.

To conclude, since $\mu'(H^\oplus) = 1$, thus the restriction of $\mu'$ to $\mathcal{A}^\oplus$ is a member of $\mathbb{P}^\sigma_{\mathsf{na}}(\mathcal{A}^\oplus)$ that represents $(\mathcal{A}^\oplus, \supseteq^\oplus, \succsim^\oplus)$, so by our earlier application of \hyperlink{TheoremV2}{Theorem~V2} this must be $\mu^\oplus$. Similarly, if we define $s' \equiv \mu'(H^\ominus)$, then the restriction of $-\frac{1}{s'}\mu'$ to $\mathcal{A}^\ominus$ is a member of $\mathbb{P}^\sigma_{\mathsf{na}}(\mathcal{A}^\ominus)$ that represents $(\mathcal{A}^\ominus, \supseteq^\ominus, \succsim^\ominus)$, so by our earlier application of \hyperlink{TheoremV2}{Theorem~V2} this must be $\mu^\ominus$. Finally, as in Step~2, there is $A \in \mathcal{A}$ such that $A$ is an annulment of $H^\ominus$ contained in $H^\oplus$, and since $A \cup H^\ominus \sim \emptyset$ we have $\mu'(A) + \mu'(H^\ominus) = \mu'(A \cup H^\ominus) = \mu'(\emptyset) = 0$, so we must have $s' = \mu'(H^\ominus) = - \mu'(A)$. Altogether, then, $\mu'$ satisfies all the properties required by our construction in Step~2, so $\mu' = \mu$, as desired.~$\blacksquare$

\hypertarget{AppendixG}{}
\section{Appendix G - Bayesian representation}

In this appendix, we prove \hyperlink{Theorem2}{Theorem~2}.

\vspace{\baselineskip} \noindent \textsc{Theorem 2:} A {\it non-degenerate} inference is smooth and {\it absolute} if and only if it has a normalized representation $\mu \in \mathbb{S}^\sigma_{\mathsf{na}}(\mathcal{A})$ such that $\mu(S) = 0$. In this case, $\mu$ is the unique normalized signed measure representation, there is a Hahn decomposition, and $\mu$ has a unique Jordan decomposition $(\mu^J_0, \mu^J_1) \in \mathbb{P}^\sigma_{\mathsf{na}} \times \mathbb{P}^\sigma_{\mathsf{na}}$. In particular, for each Hahn decomposition $(H^\oplus, H^\ominus)$ and each $A \in \mathcal{A}$,
\begin{align*}
\mu^J_0(A) &= -\inf \{\mu(B) | B \in \mathcal{A} \text{ and } B \subseteq A\} = -\mu(A \cap H^\ominus), \text{ and}
\\ \mu^J_1(A) &= \sup \{\mu(B) | B \in \mathcal{A} \text{ and } B \subseteq A\} = \mu(A \cap H^\oplus).
\end{align*}
Moreover, for each Hahn decomposition $(H^\oplus, H^\ominus)$, a tuple $(\mu_0, \mu_1, A^*) \in \mathbb{P}^\sigma_{\mathsf{na}} \times \mathbb{P}^\sigma_{\mathsf{na}} \times \mathcal{A}$ is a Bayesian representation if and only if
\begin{itemize}
\item $\mu_0 = (1 - \mu_0(H^\oplus)) \cdot \mu^J_0 + \mu_0(H^\oplus) \cdot \mu^J_1$,

\item $\mu_1 = \mu^J_1$,

\item $\mu_0(H^\oplus) \in (0, 1)$, and

\item $H^\oplus$ is equivalent to $A^*$ in the following senses: $A^* \sim H^\oplus$, $\mu_0(A^*) = \mu_0(H^\oplus)$, and $\mu_1(A^*) = \mu_1(H^\oplus)$.
\end{itemize}
Thus across all Bayesian representations, the posterior is unique, all clue guesses are in a suitable sense equivalent, and the prior is determined by the weight it assigns to each possible guess.

\vspace{\baselineskip} \noindent \textsc{Proof:} The proof consists of two steps.

\vspace{\baselineskip} \noindent $\circ$ \textsc{Step 1:} Establish the first three sentences of the theorem.

\vspace{\baselineskip} The theorem's first sentence is a direct corollary of \hyperlink{Theorem1}{Theorem~1}; thus let the inference and $\mu$ satisfy these conditions. By \hyperlink{Proposition1}{Proposition~1}, $\mu$ is the unique normalized signed measure representation, and by \hyperlink{Proposition5}{Proposition~5}, there is a Hahn decomposition $(H^\oplus, H^\ominus)$.

The existence and uniqueness of the Jordan decomposition of $\mu$ follows from standard arguments and simple observations. Indeed, let $(H^\oplus, H^\ominus)$ be a Hahn decomposition, and for each $A \in \mathcal{A}$, define $\mu_0(A) \equiv -\mu(A \cap H^\ominus)$ and $\mu_1(A) \equiv \mu(A \cap H^\oplus)$. Clearly, (i)~we have both equalities in theorem's third sentence for the given Hahn decomposition, (ii)~$\mu = \mu^J_1 - \mu^J_0$, and (iii)~both $\mu^J_0$ and $\mu^J_1$ are non-negative. Moreover, by {\it absoluteness} we have $\mu(S) = 0$, so since $\mu$ is normalized we have $\mu(H^\ominus) = -1$ and $\mu(H^\oplus) = 1$; thus $\mu^J_0(S) = 1$ and $\mu^J_1(S) = 1$. Finally, (i)~by the proof of 326L in Chapter~32 of \cite{Fremlin2012}, both $\mu^J_0$ and $\mu^J_1$ are countably additive, and (ii)~since $\mu$ has no measure-atoms, thus neither $\mu^J_0$ nor $\mu^J_0$ has measure-atoms. Altogether, then, $(\mu^J_0, \mu^J_1)$ is a Jordan decomposition of $\mu$, and by the elementary argument of \cite{Fischer2024}, there is no other Jordan decomposition of $\mu$. We therefore have the theorem's second sentence, and since $(H^\oplus, H^\ominus)$ was an arbitrary Hahn decomposition, we have the theorem's third sentence.~$\square$

\vspace{\baselineskip} \noindent $\circ$ \textsc{Step 2:} Establish the final sentence of the theorem.

\vspace{\baselineskip} Let $(H^\oplus, H^\ominus)$ be a Hahn decomposition and let $(\mu_0, \mu_1, A^*) \in \mathbb{P}^\sigma_{\mathsf{na}} \times \mathbb{P}^\sigma_{\mathsf{na}} \times \mathcal{A}$. It is straightforward to show that if $(\mu_0, \mu_1, A^*)$ satisfies the four conditions, then it is a Bayesian representation; we omit the argument. Thus let us assume that $(\mu_0, \mu_1, A^*)$ is a Bayesian representation and define $\mu_{\Delta} \equiv \mu_1 - \mu_0$. Then (i)~$\mu_\Delta$ is a signed measure representation, (ii)~$\mu_0(A^*) \in (0, 1)$, and (iii)~for each $A \in \mathcal{A}$, $\mu_1(A) = \frac{\mu_0(A \cap A^*)}{\mu_0(A^*)}$.

Let $(\mathcal{A}^\oplus, \supseteq^\oplus, \succsim^\oplus)$ and $(\mathcal{A}^\ominus, \supseteq^\ominus, \succsim^\ominus)$ denote the $H^\oplus$-subspace and the $H^\ominus$-subspace, respectively. By \hyperlink{Lemma11}{Lemma~11}, there is a unique $\mu^\oplus \in \mathbb{P}^\sigma_{\mathsf{na}}(\mathcal{A}^\oplus)$ such that $\mu^\oplus$ represents the former and there is a unique $\mu^\ominus \in \mathbb{P}^\sigma_{\mathsf{na}}(\mathcal{A}^\ominus)$ such that $-\mu^\ominus$ represents the latter. As argued in the previous step, $\mu(H^\oplus) = 1$ and $\mu(H^\ominus) = -1$; it follows that $\mu^\oplus$ is the restriction of $\mu$ to $\mathcal{A}^\oplus$ and $\mu^\ominus$ is the restriction of $\mu$ to $\mathcal{A}^\ominus$.

First, we claim that for each $A \in \mathcal{A}$ such that $A \subseteq H^\oplus \backslash A^*$, we have $\mu_0(A) = \mu_1(A) = \mu_\Delta(A) = 0$. Indeed, since $A \subseteq H^\oplus$, thus $A \succsim \emptyset$, so since $\mu_\Delta$ is a representation we have $\mu_\Delta(A) \geq 0$. Moreover, since $A \cap A^* = \emptyset$, thus $\mu_1(A) = 0$, so $\mu_\Delta(A) = -\mu_0(A) \leq 0$. Altogether, then, $\mu_1(A) = \mu_\Delta(A) = 0$, so $\mu_0(A) = 0$.

Second, we claim that for each $A \in \mathcal{A}$ such that $A \subseteq A^* \backslash H^\oplus$, we have $\mu_0(A) = \mu_1(A) = \mu_\Delta(A) = 0$.  Indeed, since $A \subseteq S \backslash H^\oplus = H^\ominus$, thus $A \precsim \emptyset$, so since $\mu_\Delta$ is a representation we have $\mu_\Delta(A) \leq 0$. Moreover, since $A \cap A^* = A$ and $\mu_0(A^*) \in (0, 1]$, thus $\mu_1(A) = \frac{1}{\mu_0(A^*)} \cdot \mu_0(A) \geq \mu_0(A)$, so $\mu_\Delta(A) = \mu_1(A) - \mu_0(A) \geq 0$. Then $\mu(A^*) \in (0, 1]$ and $\frac{1}{\mu(A^*)} \cdot \mu_0(A) = \mu_0(A)$, so $\mu_0(A) = 0$. Altogether, then, $\mu_0(A) = \mu_\Delta(A) = 0$, so $\mu_1(A) = 0$.

Third, we claim that $\mu_0(H^\oplus) = \mu_0(A^*)$, $\mu_1(H^\oplus) = \mu_1(A^*)$, $\mu_\Delta(H^\oplus) = \mu_\Delta(A^*)$, and $H^\oplus \sim A^*$. Indeed, the three equalities follow from the first two claims, and the comparison follows because $\mu_\Delta$ is a representation.

Fourth, we claim that for each $A \in \mathcal{A}$, $\mu_1(A) = \frac{\mu_0(A \cap H^\oplus)}{\mu_0(H^\oplus)}$. Indeed, let $A \in \mathcal{A}$. By the first two claims, $\mu_0((A \cap H^\oplus) \backslash A^*) = \mu_0((A \cap A^*) \backslash H^\oplus) = 0$ and $\mu_0(H^\oplus \backslash A^*) = \mu_0(A^* \backslash H^\oplus) = 0$, so $\mu_0(A \cap H^\oplus) = \mu_0(A \cap H^\oplus \cap A^*) = \mu_0(A \cap A^*)$ and $\mu_0(H^\oplus) = \mu_0(H^\oplus \cap A^*) = \mu_0(A^*)$, so $\mu_1(A) = \frac{\mu_0(A \cap A^*)}{\mu_0(A^*)} = \frac{\mu_0(A \cap H^\oplus)}{\mu_0(H^\oplus)}$, as desired.

Fifth, we claim that for each $A \in \mathcal{A}^\oplus$, $\mu_1(A) = \mu^J_1(A)$. Indeed, for each $A \in \mathcal{A}^\oplus$, by the previous two claims we have $\mu_1(A) = \frac{\mu_0(A \cap H^\oplus)}{\mu_0(H^\oplus)} = \frac{1}{\mu_0(H^\oplus)} \cdot \mu_0(A) = \frac{1}{\mu_0(A^*)} \cdot \mu_0(A)$. Then for each pair $A, B \in \mathcal{A}^\oplus$, (i)~since $\mu_\Delta$ is a representation, we have $A \succsim B$ if and only if $\mu_\Delta(A) \geq \mu_\Delta(B)$; (ii)~by the previous sentence, this is true if and only if $\frac{1 - \mu_0(A^*)}{\mu_0(A^*)} \cdot \mu_0(A) \geq \frac{1 - \mu_0(A^*)}{\mu_0(A^*)} \cdot \mu_0(B)$; and (iii)~since $\mu_0(A^*) \in (0, 1)$, this is true if and only if $\mu_1(A) = \frac{1}{\mu_0(A^*)} \cdot \mu_0(A) \geq \frac{1}{\mu_0(A^*)} \cdot \mu_0(B) = \mu_1(B)$. Then by Step~2, the restriction of $\mu_1$ to $\mathcal{A}^\oplus$ is equal to $\mu^\oplus$, so for each $A \in \mathcal{A}^\oplus$ we have $\mu_1(A) = \mu^\oplus(A) = \mu^J_1(A)$, as desired.

Sixth, we claim that $\mu_1 = \mu^J_1$. Indeed, using the fourth claim, for each $A \in \mathcal{A}^\ominus$ we have $\mu^J_1(A) = \frac{\mu(A \cap H^\oplus)}{\mu(H^\oplus)} = 0$ and $\mu_1(A) = \frac{\mu_0(A \cap H^\oplus)}{\mu_0(H^\oplus)} = 0$. Then by the previous claim, for each $A \in \mathcal{A}$, we have $\mu_1(A) = \mu_1(A \cap H^\oplus) + \mu_1(A \backslash H^\oplus) = \mu^J_1(A \cap H^\oplus) + \mu^J_1(A \backslash H^\oplus) = \mu^J_1(A)$, as desired.

Finally, we conclude. Indeed, for each $A \in \mathcal{A}$, since $\mu_\Delta$ is a representation and {\it absoluteness} is satisfied, thus $\mu_\Delta(H^\oplus) \geq \mu_\Delta(A) \geq \mu_\Delta(H^\ominus) = - \mu_\Delta(H^\oplus)$, so $\frac{1}{\mu_\Delta(H^\oplus)} \cdot \mu_\Delta$ is a normalized representation, so by \hyperlink{Theorem1}{Theorem~1} it is $\mu$. Then by the fourth claim, the sixth claim, and the third claim, we have
\begin{align*}
\mu^J_1 - \mu^J_0 & = \mu
\\ &= \frac{1}{\mu_\Delta(H^\oplus)} \cdot \mu_\Delta
\\ & = \frac{1}{1 - \mu_0(H^\oplus)} \cdot (\mu_1 - \mu_0)
\\ & = \frac{1}{1 - \mu_0(H^\oplus)} \cdot (\mu^J_1 - \mu_0)
\\ & = \frac{1}{1 - \mu_0(A^*)} \cdot (\mu^J_1 - \mu_0),
\end{align*}
so $(1 - \mu_0(A^*)) \cdot (\mu^J_1 - \mu^J_0) = \mu^J_1 - \mu_0$, or $\mu_0 = (1 - \mu_0(A^*)) \cdot \mu^J_0 + \mu_0(A^*) \cdot \mu^J_1$, as desired.~$\blacksquare$

\phantomsection


\begin{thebibliography}{}

{\footnotesize

\bibitem[\protect\citeauthoryear{Anscombe and Aumann}{1963}]{Anscombe-Aumann1963}
\textsc{Anscombe, Frank and Robert Aumann} (1963). ``A definition of subjective probability." {\it The Annals of Mathematical Statistics} 34, 199-205.

\bibitem[\protect\citeauthoryear{Arrow}{1964}]{Arrow1964}
\textsc{Arrow, Kenneth} (1964). ``The role of securities in the optimal allocation of risk-bearing." {\it The Review of Economic Studies} 31, 91-96.

\bibitem[\protect\citeauthoryear{Arrow}{1970}]{Arrow1970}
\textsc{Arrow, Kenneth} (1970). {\it Essays in the Theory of Risk-Bearing}. Amsterdam, the Netherlands: North-Holland Publishing Company.

\bibitem[\protect\citeauthoryear{Balcar and Jech}{2006}]{Balcar-Jech2006}
\textsc{Balcar, Bohuslav and Thomas Jech} (2006). ``Weak distributivity, a problem of von Neumann and the mystery of measurability." {\it The Bulletin of Symbolic Logic} 12, 241-266.

\bibitem[\protect\citeauthoryear{Balcar, Jech, and Paz\'{a}k}{2005}]{Balcar-Jech-Pazak2005}
\textsc{Balcar, Bohuslav, Thomas Jech, and Tom\'{a}\v{s} Paz\'{a}k} (2005). ``Complete ccc Boolean algebras, the order sequential topology, and a problem of von Neumann." {\it Bulletin of the London Mathematical Society} 37, 885-898.

\bibitem[\protect\citeauthoryear{Bernstein}{1917}]{Bernstein1917}
\textsc{Bernstein, Sergei} (1917). ``Attempt at an axiomatic foundation of probability theory [in Russian]." {\it Communications of the Kharkov Mathematical Society} [in Russian] 15, 209-274. Translation In: Sheynin,~O. (2005). {\it Probability and Statistics: Russian Papers of the Soviet Period}. Berlin, Germany: NG Verlag.

\bibitem[\protect\citeauthoryear{Bewley}{2002}]{Bewley2002}
\textsc{Bewley, Truman} (2002). ``Knightian decision theory. Part I." {\it Decisions in Economics and Finance} 25, 79-110.

\bibitem[\protect\citeauthoryear{Brandenburger, Ghirardato, Pennesi, and Stanca}{2024}]{Brandenburger-Ghirardato-Pennesi-Stanca2024}
\textsc{Brandenburger, Adam, Paolo Ghirardato, Daniele Pennesi, and Lorenzo Stanca} (2024). ``Event valence and subjective probability." Working paper.

\bibitem[\protect\citeauthoryear{Dominiak, Kovach, and Tserenjigmid}{2025}]{Dominiak-Kovach-Tserenjigmid2025}
\textsc{Dominiak, Adam, Matthew Kovach, and Gerelt Tserenjigmid} (2025). ``Inertial updating with general information." Working paper.

\bibitem[\protect\citeauthoryear{Dybvig and Ross}{2003}]{Dybvig-Ross2003}
\textsc{Dybvig, Philip and Stephen Ross} (2003). ``Arbitrage, state prices and portfolio theory." {\it Handbook of the Economics of Finance, Volume 1}, 605-637.

\bibitem[\protect\citeauthoryear{Fama}{1970}]{Fama1970}
\textsc{Fama, Eugene} (1970). ``Efficient capital markets: a review of theory and empirical work." {\it The Journal of Finance} 25, 383-417.

\bibitem[\protect\citeauthoryear{Fremlin}{2012}]{Fremlin2012}
\textsc{Fremlin, David} (2012). {\it Measure Theory, Volume 3, Part 1 (Second Edition).} Self-published, Lulu.

\bibitem[\protect\citeauthoryear{de Finetti}{1937}]{deFinetti1937}
\textsc{de Finetti, Bruno} (1937). ``La pr\'{e}vision: ses lois logiques, ses sources subjectives [in French]." {\it Annales d'Institut Henri Poincar\'{e}} 7, 1-68.

\bibitem[\protect\citeauthoryear{Fischer}{2024}]{Fischer2024}
\textsc{Fischer, Tom} (2024). ``Existence, uniqueness, and minimality of the Jordan measure decomposition." Working paper.

\bibitem[\protect\citeauthoryear{Hahn}{1921}]{Hahn1921}
\textsc{Hahn, Hans} (1921). {\it Theorie der reellen Funktionen} [in German]. Berlin, Germany: Verlag von Julius Springer.

\bibitem[\protect\citeauthoryear{Halmos}{1963}]{Halmos1963}
\textsc{Halmos, Paul} (1963). {\it Lectures on Boolean Algebras.} New York, NY: Van Nostrand.

\bibitem[\protect\citeauthoryear{Jeffrey}{1965}]{Jeffrey1965}
\textsc{Jeffrey, Richard} (1965). {\it The Logic of Decision.} New York, NY: McGraw-Hill.

\bibitem[\protect\citeauthoryear{Jordan}{1893}]{Jordan1893}
\textsc{Jordan, Camille} (1893). {\it Cours d'analyse de l'\'{E}cole polytechnique} [in French]. Paris, France: Gauthier-Villars et fils.

\bibitem[\protect\citeauthoryear{Koopman}{1940}]{Koopman1940}
\textsc{Koopman, Bernard} (1940). ``The axioms and algebra of intuitive probability." {\it Annals of Mathematics} 41, 269-292.

\bibitem[\protect\citeauthoryear{Kopylov}{2007}]{Kopylov2007}
\textsc{Kopylov, Igor} (2007). ``Subjective probabilities on `small' domains." {\it Journal of Economic Theory} 113, 236-265.

\bibitem[\protect\citeauthoryear{Kraft, Pratt, and Seidenberg}{1959}]{Kraft-Pratt-Seidenberg1959}
\textsc{Kraft, Charles, John Pratt and Abraham Seidenberg} (1959). ``Intuitive probability on finite sets." {\it Annals of Mathematical Statistics} 30, 408-419.

\bibitem[\protect\citeauthoryear{Loomis}{1947}]{Loomis1947}
\textsc{Loomis, Lynn} (1947). ``On the representation of $\sigma$-complete Boolean algebras." {\it Bulletin of the American Mathematical Society} 53, 757-760.

\bibitem[\protect\citeauthoryear{Maharam}{1947}]{Maharam1947}
\textsc{Maharam, Dorothy} (1947). ``An algebraic characterization of measure algebras." {\it Annals of Mathematics} 48, 154-167.

\bibitem[\protect\citeauthoryear{Mackenzie}{2019}]{Mackenzie2019}
\textsc{Mackenzie, Andrew} (2019). ``A foundation for probabilistic beliefs with or without atoms." {\it Theoretical Economics} 14, 709-778.

\bibitem[\protect\citeauthoryear{Nikodym}{1930}]{Nikodym1930}
\textsc{Nikodym, Otton} (1930). ``Sur une g\'{e}n\'{e}ralisation des int\'{e}grales de M. J. Radon [in French]." {\it Fundamenta Mathematicae} 15, 131-179.

\bibitem[\protect\citeauthoryear{Ok, Ortoleva, and Riella}{2012}]{Ok-Ortoleva-Riella2012}
\textsc{Ok, Efe, Pietro Ortoleva, and Gil Riella} (2012). ``Incomplete preferences under uncertainty: indecisiveness in beliefs versus tastes." {\it Econometrica} 80, 1791-1808.

\bibitem[\protect\citeauthoryear{Radon}{1913}]{Radon1913}
\textsc{Radon, Johann} (1913). ``Theorie und Anwendungen der absolut additiven Mengenfunktionen [in German]." {\it Sitzungsberichte der Kaiserlichen Akademie der Wissenschaften: Mathematisch-Naturwissenschaftliche Klasse} 122, Issue 1, Part IIa, 1295-1438.

\bibitem[\protect\citeauthoryear{Ramsey}{1931}]{Ramsey1931}
\textsc{Ramsey, Frank} (1931). ``Truth and Probability." In: {\it Foundations of Mathematics and other Logical Essays.} Editor: Braithwaite, Richard. London: Kegan Paul, Trench, Trubner, and Co., Ltd.

\bibitem[\protect\citeauthoryear{Savage}{1954}]{Savage1954}
\textsc{Savage, Leonard} (1954). {\it The Foundations of Statistics.} Wiley, NY: Dover Publications.

\bibitem[\protect\citeauthoryear{Savage}{1972}]{Savage1972}
\textsc{Savage, Leonard} (1972). {\it The Foundations of Statistics: Second Revised Edition.} Wiley, NY: Dover Publications.

\bibitem[\protect\citeauthoryear{Sierpi\'{n}ski}{1922}]{Sierpinski1922}
\textsc{Sierpi\'{n}ski, Wac\l{}aw} (1922). ``Sur les fonctions d'ensemble additives et continues" [in French]. {\it Fundamenta Mathematicae} 3, 240-246.

\bibitem[\protect\citeauthoryear{Sikorski}{1960}]{Sikorski1960}
\textsc{Sikorski, Roman} (1960). {\it Boolean Algebras.} Berlin, Germany: Springer-Verlag.

\bibitem[\protect\citeauthoryear{Steinhaus}{1948}]{Steinhaus1948}
\textsc{Steinhaus, Hugo} (1948). ``The problem of fair division." {\it Econometrica} 16, 101-104.

\bibitem[\protect\citeauthoryear{Stone}{1936}]{Stone1936}
\textsc{Stone, Marshall} (1936). ``The theory of representations of Boolean algebras." {\it Transactions of the American Mathematical Society} 40, 37-111.

\bibitem[\protect\citeauthoryear{Velickov}{2005}]{Velickov2005}
\textsc{Velickov, Boban} (2005). ``CCC forcing and splitting reals." {\it Israel Journal of Mathematics} 147, 209-220.

\bibitem[\protect\citeauthoryear{Villegas}{1964}]{Villegas1964}
\textsc{Villegas, Cesareo} (1964). ``On quantitative probability $\sigma$-algebras." {\it Annals of Mathematical Statistics} 35, 1787-1796.

\bibitem[\protect\citeauthoryear{Vladimirov}{2002}]{Vladimirov2002}
\textsc{Vladimirov, Denis} (2002). {\it Boolean Algebras in Analysis.} Dodrecht, the Netherlands: Kluwer Academic Publishers.

\bibitem[\protect\citeauthoryear{Zorn}{1935}]{Zorn1935}
\textsc{Zorn, Max} (1935). ``A remark on method in transfinite algebra." {\it Bulletin of the American Mathematical Society} 41, 667-670.

}

\end{thebibliography}
\end{document}